\documentclass[twocolumn,twocolappendix]{aastex631}

\defcitealias{2019ApJ...876...27Y}{YL19}
\defcitealias{2021ApJ...919L..16Y}{YLM21}

\shorttitle{Resurrection of Type Ib/c SNRs}
\shortauthors{Yasuda et al.}
\graphicspath{{./}{figures/}}

\begin{document}

\title{Resurrection of Non-thermal Emissions from Type Ib/c Supernova Remnants}

\author[0000-0002-0802-6390]{Haruo Yasuda}
\affiliation{Department of Astronomy, Kyoto University, Kitashirakawa, Oiwake-cho, Sakyo-ku, Kyoto 606-8502, Japan}

\author[0000-0002-2899-4241]{Shiu-Hang Lee}
\affiliation{Department of Astronomy, Kyoto University, Kitashirakawa, Oiwake-cho, Sakyo-ku, Kyoto 606-8502, Japan}
\affiliation{Kavli Institute for the Physics and Mathematics of the Universe (WPI), The University of Tokyo, Kashiwa 277-8583, Japan}

\author[0000-0003-2611-7269]{Keiichi Maeda}
\affiliation{Department of Astronomy, Kyoto University, Kitashirakawa, Oiwake-cho, Sakyo-ku, Kyoto 606-8502, Japan}

\correspondingauthor{Haruo Yasuda}
\email{yasuda@kusastro.kyoto-u.ac.jp}



\begin{abstract}
Supernova remnants (SNRs) are important objects in investigating the links among supernova (SN) explosion mechanism(s), progenitor stars, and cosmic-ray acceleration. Non-thermal emission from SNRs is an effective and promising tool for probing their surrounding circumstellar media (CSM) and, in turn, the stellar evolution and mass-loss mechanism(s) of massive stars. In this work, we calculate the time evolution of broadband non-thermal emissions from Type Ib/c SNRs whose CSM structures are derived from the mass-loss history of their progenitors. Our results predict that Type Ib/c SNRs make a transition of brightness in radio and $\gamma$-ray bands from an undetectable dark for a certain period to a re-brightening phase. This  transition originates from their inhomogeneous CSM structures in which the SNRs are embedded within a low-density wind cavity surrounded by a high-density wind shell and the ambient interstellar medium (ISM). The ``resurrection'' in non-thermal luminosity happens at an age of $\sim$1,000 yrs old for a Wolf-Rayet star progenitor evolved within a typical ISM density. Combining with the results of Type II SNR evolution recently reported by \cite{2021ApJ...919L..16Y}, this result sheds light on a comprehensive understanding of non-thermal emissions from SNRs with different SN progenitor types and ages, which is made possible for the first time by the incorporation of realistic mass-loss histories of the progenitors.

\end{abstract}



\section{Introduction}\label{sec:Intro}

Core-collapse supernovae (SNe) are classically divided into two major classes; Type II SNe and Type Ib/c SNe \citep{1985ApJ...296..379E,1986ASIC..180...45W}. This classification is based on the presence or absence of absorption lines from H and He in their spectra around the maximum light. 
The difference is believed to be originated from the differences in the nature of their progenitor stars and the associated mass-loss histories. The classification of SN types is hence important for the investigation of stellar evolution of massive stars and their explosion mechanism(s). 

The rate of Type Ib/c SNe is estimated to be about one third of their Type II counterpart \citep[e.g.,][]{2011MNRAS.412.1522S}. However, a smoking-gun observational evidence for a Type Ib/c supernova remnant (SNR) is still absent. Type Ib/c SNe are also noteworthy from the perspective of the production of neutron star systems and millisecond pulsars \citep[e.g.,][]{1999A&A...350..928T,2009ASSL..359..125V,2021MNRAS.506.4654W}, and theoretical studies of Type Ib/c SNe have developed rapidly in the past few decades \citep[e.g.,][]{2017hsn..book..403S,2017MNRAS.470.3970Y,2019ApJ...878...49W,2020ApJ...890...51E,2020ApJ...896...56W,2021ApJ...913..145W}. On the other hand, detailed evolution and emission models for Type Ib/c SNRs are still scarce in the literature, which however are essential for their future identifications and a comprehensive understanding of the SNR population. In fact, there are only few examples of known Galactic SNRs which are speculated to bear a Type Ib/c origin, such as RX J1713.7-3946 \citep{2015ApJ...814...29K}. A theoretical study linking SNe and SNRs \citep[][hereafter \citetalias{2021ApJ...919L..16Y}]{2021ApJ...919L..16Y} in a self-consistent evolution model is an urgent and crucial task. 

In this work, we first prepare self-consistent CSM models taking into account the stellar evolution and mass-loss history of a Type Ib/c progenitor using one-dimensional hydrodynamic simulations. Second, we calculate long-term time evolution of the SNR dynamics and the resulted non-thermal emissions produced by the interaction of the SNR with their CSM environments up to an age of $10^4$ yr. In Section~\ref{sec:Method}, we briefly introduce our simulation method for the hydrodynamics, particle acceleration, and the construction of CSM models aided by knowledge from SN observations and progenitor models. In Section~\ref{sec:Results}, we show the results on the non-thermal emissions from Ib/c SNRs with different progenitor masses and CSM structures, and their detectability by currently available and future detectors. Discussion and conclusions can be found in Section~\ref{sec:Discussion} and Section~\ref{sec:Conclusion}.

\section{Method}\label{sec:Method}

\subsection{Particle acceleration and Hydrodynamics} \label{physics}
We use the well-tested {\it CR-Hydro} hydrodynamic code as in \citetalias{2021ApJ...919L..16Y}, which has also been used recently in \citet[][hereafter \citetalias{2019ApJ...876...27Y}]{2019ApJ...876...27Y} and decades of previous works referenced therein. This code simultaneously calculates the hydrodynamic evolution of a SNR coupled to the particle acceleration at the SNR shocks and the accompanying multi-wavelength emission in a time and space-resolved fashion. In this section, we introduce the {\it CR-Hydro} code briefly, and other details can be referred in \citetalias{2019ApJ...876...27Y,2021ApJ...919L..16Y}. 

This code solves the hydrodynamic equations written in Lagrangian coordinate $m$ which assumes a spherical symmetry and includes feedbacks from efficient cosmic rays (CRs) production via nonlinear diffusive shock acceleration;
\begin{eqnarray}
&&\frac{\partial r}{\partial m} + \frac{1}{4\pi r^2\rho} = 0\\
&&\frac{\partial u}{\partial t} + \frac{\partial}{\partial m}\left(P_\mathrm{g}+P_\mathrm{CR}+\frac{B^2}{8\pi}\right) = 0\\
&&\frac{\partial e}{\partial t} + \frac{\partial }{\partial m}\left[\left(P_\mathrm{g}+P_\mathrm{CR}+\frac{B^2}{8\pi}\right)u\right] = -n^2\Lambda_\mathrm{cool}\\
&&e=\frac{1}{2}u^2+\frac{1}{\rho}\left(\frac{P_\mathrm{g}}{\gamma_\mathrm{g}-1}+\frac{P_\mathrm{CR}}{\gamma_\mathrm{CR}-1}+\frac{B^2}{8\pi}\right).
\end{eqnarray}
The code also solves the diffusion-convection equation written in the shock-rest frame assuming a steady-state and isotropic distribution of the accelerated particles in momentum space \citep{2010APh....33..307C,2010MNRAS.407.1773C,2012ApJ...750..156L};
\begin{eqnarray}\label{eq:DSA}
&&[u(x)-v_A(x)]\frac{\partial f_p(x,p)}{\partial x}-\frac{\partial}{\partial x}\left[D(x,p)\frac{\partial f_p(x,p)}{\partial x}\right] \nonumber\\
&&=\frac{p}{3}\frac{d[u(x)-v_A(x)]}{dx}\frac{\partial f_p(x,p)}{\partial p}+Q_p(x,p).
\end{eqnarray}
In the above equations, $\rho$, $P_\mathrm{g}$, and $u$ are the mass density, pressure, and flow velocity of thermal gas, $P_\mathrm{CR}=\int 4\pi p^2f_\mathrm{p}dp$ is the CR pressure, and $f_\mathrm{p}(x,p)$ is the phase-space distribution function of the accelerated protons. By adopting the so-called thermal-leakage model \citep{2004APh....21...45B,2005MNRAS.361..907B} as a convenient parameterization for the DSA injection term $Q_p(x,p)$, we can obtain the semi-analytic solution of $f_p(x,p)$ \citep{2010APh....33..307C,2010MNRAS.407.1773C,2012ApJ...750..156L}, for which the explicit expression can be found in \citetalias{2019ApJ...876...27Y,2021ApJ...919L..16Y}. The treatment of the magnetic field strength $B(x)$ and the spatial diffusion coefficient $D(x,p)$ of the accelerating particles can also be found in \citetalias{2021ApJ...919L..16Y}. 

In addition, we parametrically treat the electron distribution function as $f_e(x,p) = K_\mathrm{ep}f_p(x,p)\exp[-(p/p_\mathrm{max})^{\alpha_\mathrm{cut}}]$. $K_\mathrm{ep}$ typically takes a value between $10^{-3}$ and $10^{-2}$ which is limited by SNR observations so far. The determination of the maximum momenta $p_\mathrm{max}$ and the cut-off index $\alpha_\mathrm{cut}$ is done in the same way as in \citetalias{2019ApJ...876...27Y}. 

The accelerated particles are advected to the downstream and are assumed to be co-moving with the post-shock gas flow by magnetic confinement. Both the freshly accelerated particles at the shock and the advected particles interact with their surrounding gas to produce multi-wavelength non-thermal emissions and meanwhile lose their energies through radiation and adiabatic expansion. In our models, we include non-thermal radiation mechanisms by synchrotron radiation, inverse Compton scattering (IC), non-thermal bremsstrahlung from the accelerated electrons, and pion productions and decay from proton-proton interactions ($\pi^0$ decay) by the accelerated protons. In this study, only the cosmic microwave background radiation (CMB) is considered for the target photon field of IC for generality, but this can be modified when we target any specific SNR.

We note that our code is similar in construct to that used in some previous works \citep{2010ApJ...718...31P,2012APh....39...12Z}. \citet{2012APh....39...12Z} conducted simulations of SNR evolution and particle acceleration in detail, in which they consider models typical of Type Ia SNRs evolving in a uniform ISM-like environment. In this work, we have included a few additional physical components such as radiative cooling, a treatment of magnetic field amplification (MFA) from resonant streaming instability \citep{1978MNRAS.182..147B,1978MNRAS.182..443B}, as well as a spatially inhomogeneous CSM environment for core-collapse SNRs motivated by the time-dependent mass loss histories of their progenitors prior to explosion, as we will discuss in more detail in the next section.

\subsection{Circumstellar medium and SN ejecta}\label{Method:CSM}
In this study, we first construct CSM models for a Type Ib/c SNR by performing hydrodynamic simulations in which stellar winds from the progenitor run into a uniform ISM. We account for the stellar evolution and mass-loss histories of the SN progenitor under a grid of model parameters inspired by observations. These results are used as the initial conditions for calculating the subsequent long-term evolution of the SNR.  

The progenitor of a Type Ib/c SN is usually linked to massive OB-type stars with zero-age main sequence (ZAMS) mass $\ge10M_\odot$ in a binary system.
When the progenitors evolve to red supergiants (RSG) after their main sequence (MS), their envelopes fill the Roche-lobe and the hydrogen envelopes are stripped by a Roche-lobe overflow (RLOF). As a result, they evolve to a helium or carbon-oxygen star called a Wolf-Rayet (WR) star, which eventually explode via core collapse. The stellar wind blown in the MS and WR phases are fast because of the compactness of the OB and WR stars, and the total amount of mass lost in these phases is relatively small. On the other hand, the mass-loss mechanism in the RLOF phase is still under discussion. Two channels can be considered: (i) the material stripped by RLOF is spread out into the circumstellar environment in the form of a stellar wind, and (ii) the stripped gas accretes onto the companion stars. It strongly depends on the binary properties such as the mass ratio and separation of the two stars. For simplicity, we treat the accretion efficiency as a parameter ($\beta_\mathrm{acc}$) in our models, so that $\beta_\mathrm{acc} \equiv \dot{M}_\mathrm{sec}/\dot{M}_\mathrm{pri}$ where $\dot{M}_\mathrm{pri}$ and  $\dot{M}_\mathrm{sec}$ are the mass loss rates of the donor star and the accretion rate onto the secondary star, respectively. This procedure is known as a wind Roche-lobe overflow model \citep{2012BaltA..21...88M,2013A&A...552A..26A,2019MNRAS.485.5468I}. An effective mass loss rate is then obtained as $(1-\beta_\mathrm{acc})\dot{M}_\mathrm{pri}$. In this paper, we consider two extreme cases of $\beta_\mathrm{acc}=0$ and $\beta_\mathrm{acc}=1$, and adopt the $\beta_\mathrm{acc}=0$ case for our main results. Results from the $\beta_\mathrm{acc}=1$ case is discussed in Appendix~\ref{Appendix} for reference. 

The evolution of helium stars up to core-collapse is well studied by simulations \citep[e.g.,][]{2017MNRAS.470.3970Y,2019ApJ...878...49W,2020ApJ...890...51E,2021ApJ...913..145W,2021ApJ...916L...5V}, from which the mass lost in each evolutionary phase and the ejecta mass $M_\mathrm{ej}$ can be determined for a given ZAMS mass. In this study, we consider two cases for the ZAMS mass in our fiducial models, i.e., a $12M_\odot$ (model A) and $18M_\odot$ (model B) progenitor star. For comparison, we prepare two additional models (model C and D) in which the ZAMS mass is the same but the mass loss in the MS stage is not taken into account. 

We note that there is another possible way for the stars to explode as Type Ib/c SNe. A star more massive than $M_\mathrm{ZAMS}\ge30\ M_\odot$ may evolve as a single star from MS to RSG and becomes a WR star if its mass-loss rate is high enough to strip off their entire hydrogen envelope in the RSG phase. These stars also have massive helium cores, so in order to explode as Type Ib/c SNe, implying a high mass-loss rate in the WR phase as well \citep{2017MNRAS.470.3970Y,2019ApJ...878...49W,2020ApJ...890...51E,2021ApJ...913..145W}. We will discuss the results of a single star evolution model with $M_\mathrm{ZAMS}=30\ M_\odot$ in Appendix~\ref{Appendix}.

The typical mass-loss rate $\dot{M}_\mathrm{w}$, wind velocity $V_\mathrm{w}$ and time duration $\tau_\mathrm{phase}$ in each mass loss phase are $\dot{M}_\mathrm{w}\sim 10^{-8}-10^{-7}\ M_\odot\ \mathrm{yr^{-1}}$, $V_\mathrm{w}\sim1-3\times10^3\ \mathrm{km\ s^{-1}}$, and $\tau_\mathrm{phase}\sim 10^6-10^7\ \mathrm{yr}$ for the MS phase, $\dot{M}_\mathrm{w}\sim 10^{-3}-10^{-2}\ M_\odot\ \mathrm{yr^{-1}}$, $V_\mathrm{w}\sim10-100\ \mathrm{km\ s^{-1}}$, and $\tau_\mathrm{phase}\sim 10^3-10^4\ \mathrm{yr}$ for the RLOF phase, and $\dot{M}_\mathrm{w}\sim 10^{-6}-10^{-5}\ M_\odot\ \mathrm{yr^{-1}}$, $V_\mathrm{w}\sim1-3\times10^3\ \mathrm{km\ s^{-1}}$, and $\tau_\mathrm{phase}\sim 10^5-10^6\ \mathrm{yr}$ for the WR phase \citep[e.g.,][]{2017hsn..book..403S,2017MNRAS.470.3970Y,2019ApJ...878...49W,2020ApJ...890...51E,2021ApJ...913..145W}. We use a time-independent, constant mass loss rate and wind velocity during each phase for simplicity. The exact values used in the models are summarized in Table~\ref{table:wind}.

The SN ejecta mass in each model is calculated as $M_\mathrm{ej} = M_\mathrm{ZAMS} - \sum (\dot{M}_\mathrm{w}\tau_\mathrm{phase}) - M_\mathrm{rm}$, where $M_\mathrm{rm}$ is the compact remnant mass after explosion. For the ZAMS mass range we consider in this work, $M_\mathrm{rm}$ is typically $1.3\sim1.6\ M_\odot$ \citep[]{2020ApJ...896...56W}. $M_\mathrm{rm}=1.5\ M_\odot$ is adopted in all models here.
For the SN ejecta structure, we use the power-law envelope model in \citet[]{1999ApJS..120..299T} for all of our models:
\begin{equation}\label{eq:ejecta}
\rho(r)= \left \{
\begin{array}{l}
\rho_\mathrm{c} \ \ \ \ \ \ \ \ \ \ \ \ \ \ \ \  (r\le r_\mathrm{c}) \\
\rho_\mathrm{c}(r/r_\mathrm{ej})^{-n_\mathrm{SN}}\ (r_\mathrm{c}\le r \le r_\mathrm{ej}), 
\end{array}\right.
\end{equation} 
where $\rho_\mathrm{c}$, $r_\mathrm{c}$, and $r_\mathrm{ej}$ are the core density, core radius and ejecta size, respectively, which can be obtained by mass and energy conservation. We assume an explosion kinetic energey $E_\mathrm{SN} = 1.2\times10^{51}\ \mathrm{erg}$ and the power-law index of the envelope $n_\mathrm{SN} = 10$ \citep[e.g.,][]{1999ApJ...510..379M,2006ApJ...651..381C}. The ejecta masses in each model are summarized in Table~\ref{table:wind}.

\begin{deluxetable*}{ccccccccc}
\tablecaption{Model parameters \label{table:wind}}
\tablewidth{0pt}
\tablehead{
\colhead{Model} & \colhead{$M_\mathrm{ZAMS}$} & \colhead{Wind Phases} & \colhead{$\dot{M}_\mathrm{w}$} & \colhead{$V_\mathrm{w}$} & \colhead{$M_\mathrm{w}$} & \colhead{$\tau_\mathrm{phase}$} & \colhead{$M_\mathrm{ej}$}\\
\colhead{} & \colhead{($M_\odot$)} & \colhead{} & \colhead{($M_\odot\ \mathrm{yr^{-1}}$)} & \colhead{($\mathrm{km\ s^{-1}}$)} & \colhead{($M_\odot$)} & \colhead{($\mathrm{yr}$)} & \colhead{($M_\odot$)}}
\startdata
A & 12 & MS    & $5.0\times10^{-8}$ & 2000   & 0.5 & $1.0\times10^7$ &     \\
  &    & RLOF  & $8.5\times10^{-4}$ & 10     & 8.5 & $1.0\times10^4$ &     \\
  &    & WR    & $5.0\times10^{-6}$ & 2000   & 0.5 & $1.0\times10^5$ & 1.0 \\
\hline
B & 18 & MS    & $6.0\times10^{-8}$ & 2000   & 0.3 & $5.0\times10^6$ &     \\
  &    & RLOF  & $1.27\times10^{-3}$ & 10    & 12.7 & $1.0\times10^4$ &     \\
  &    & WR    & $1.0\times10^{-5}$ & 2000   & 1.0 & $1.0\times10^5$ & 2.5 \\
\hline
C & 12 & RLOF  & $9.0\times10^{-4}$ & 10     & 9.0 & $1.0\times10^4$ &     \\
  &    & WR    & $5.0\times10^{-6}$ & 2000   & 0.5 & $1.0\times10^5$ & 1.0 \\
\hline
D & 18 & RLOF  & $1.3\times10^{-3}$ & 10     & 13.0 & $1.0\times10^4$ &     \\
  &    & WR    & $1.0\times10^{-5}$ & 2000   & 1.0 & $1.0\times10^5$ & 2.5
\enddata
\tablecomments{Wind parameters and ejecta properties for a Type Ib/c SNR. The wind temperature is set to $T = 10^4\ \mathrm{K}$, SN explosion energy $E_\mathrm{SN}=1.2\times10^{51}\ \mathrm{erg}$, power-law index of the ejecta envelope $n_\mathrm{ej}=10$, and stellar remnant mass $M_\mathrm{rm} = 1.5M_\odot$ \citep[]{2020ApJ...896...56W} in all models. We also assume $n=1.0\ \mathrm{cm}^{-3}$ and $T=10^4\ \mathrm{K}$ for the outer ISM region.}
\end{deluxetable*}

\begin{figure}[ht!]
\plotone{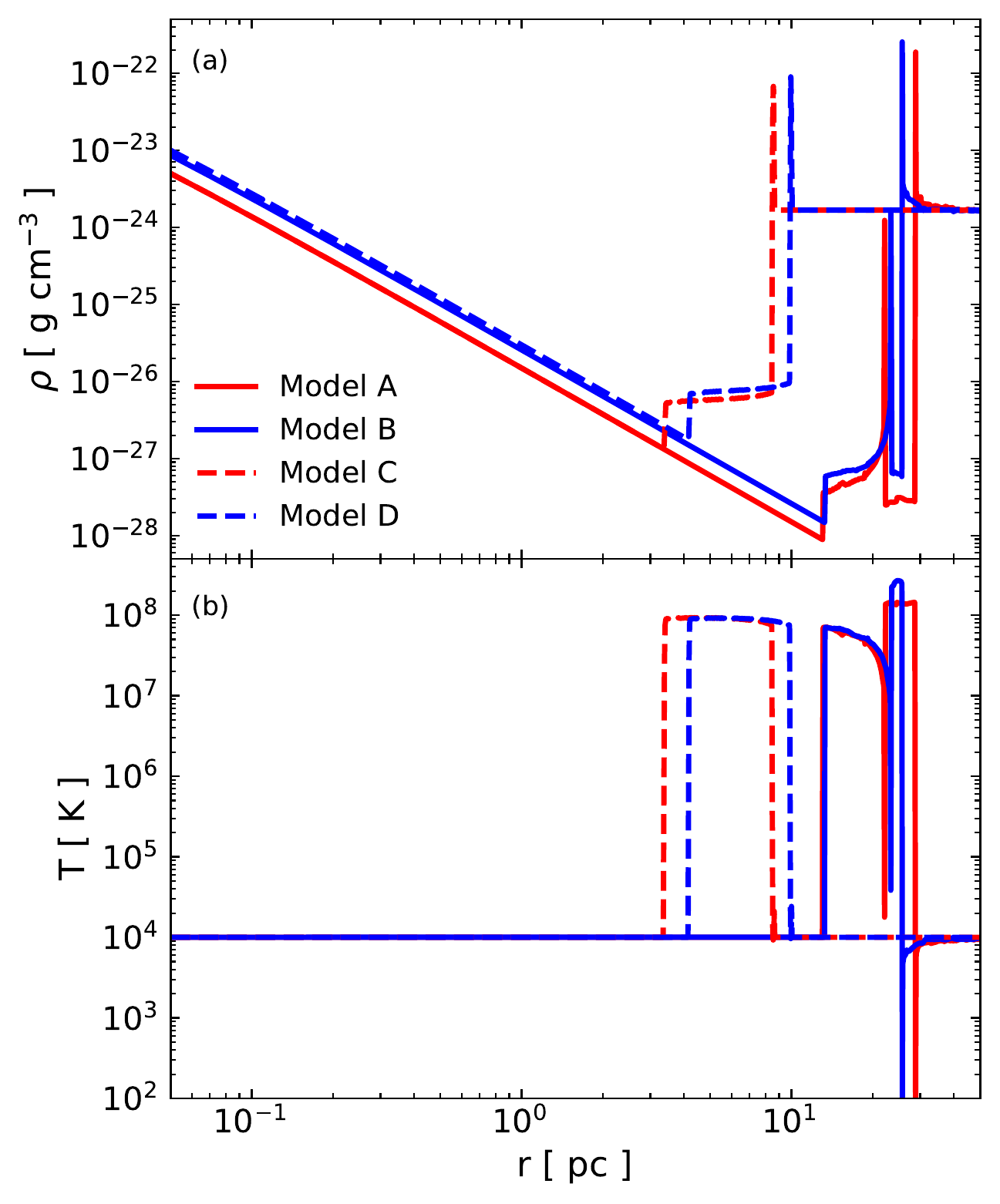}
\caption{CSM models for a Type Ib/c SNR. The upper panel shows the gas density as a function of radius, and the lower panel shows the gas temperature. The red (blue) solid line corresponds to the low (high) progenitor mass case. The dashed lines show the results from models for which the mass loss in the MS phase is not considered for comparison. \label{fig:wind}}
\end{figure}

The results of our stellar wind simulations are shown in Figure~\ref{fig:wind}. The upper panel (a) shows the radial density distribution of the CSM created by the stellar wind from a Type Ib/c SN progenitor. The lower panel (b) shows the gas temperature as a function of radius. The red and blue solid lines correspond to the results of the $12\ M_\odot$ (model A) and $18\ M_\odot$ (model B) cases in both panels, respectively. The dashed lines represent the models for which the mass loss in the MS phase is not considered for comparison (models C and D).  As the initial condition for the wind simulations, we assume a uniform ISM with $n_\mathrm{ISM}=1.0\ \mathrm{cm}^{-3}$ and $T=10^4\ \mathrm{K}$ in all of our models. 

From the results of model A and B, we can see that the CSM structure can be broken into five characteristic regions from the outer to inner radii; (i) uniform ISM, (ii) MS shell, (iii) MS bubble, (iv) WR shell, and (v) WR wind. The formation mechanism and features of the MS shell and MS bubble have been explained in detail in \citetalias{2021ApJ...919L..16Y}. Because the RLOF wind is characterized by a high mass-loss rate but slow velocity and a short time period, a dense ($\rho\ge10^{-20}\ \mathrm{g\ cm^{-3}}$) and compact ($r\le0.1\ \mathrm{pc}$) structure with a power-law profile in density is formed. On the other hand, the subsequent WR wind has a higher velocity and longer time duration than that in the RLOF phase. The fast WR wind hence sweeps up all of the above structures created by previous phases of mass loss, and creates a WR shell at $r\sim20\ \mathrm{pc}$. This result implies that the more compact structures in the CSM created before the WR phase are most probably washed away by the subsequent WR wind and accumulate onto the dense WR shell. This can also be seen in the two extra models (the $\beta_\mathrm{acc}=1$ model and the $M_\mathrm{ZAMS}=30\ M_\odot$ model single star model) to be discussed in Appendix~\ref{Appendix}. The differences in the CSM structure between models A and B are coming from the slight difference in the mass loss rate and time duration in each pre-SN evolution phase, which leads model B to have the MS shell shifted inward and the WR shell outward, and the size of MS bubble reduced compared to model A. 

On the other hand, because models C and D do not include the mass loss in the MS phase intentionally, the CSM structures in these models are relatively simple and can be divided into three main regions: (i) uniform ISM, (ii) WR shell, and (iii) WR wind. One unique feature of these models is that WR shell is located at a small radius $r\sim 10\ \mathrm{pc}$. 
In these models, the WR wind first sweeps up the dense CSM material from the RLOF phase (whose structures are almost identical to those of models A and B), beyond which the CSM density is higher than models A and B because a tenuous MS bubble is absent without the mass loss in the MS phase taken into account. The formation of the WR shell thus happens in a shorter timescale than Models A and B since the expanding WR wind cavity is sweeping up the ISM material in the downstream which has a much higher density than the tenuous MS bubble. The rapidly accumulating mass in the WR shell leads to a stronger deceleration of the cavity expansion, and hence a smaller cavity size prior to core-collapse.
There is no drastic differences between model C and D. Overall, the (non-)existence of the MS mass loss phase gives rise to the most significant variation in the hydrodynamic structure of the CSM among the models considered in this work. 

In the next step, we employ these results as the initial conditions for our simulations of the subsequent SNR evolution after explosion.  We note that we do not consider the effect of metallicity in the wind models as well as the SNR simulations, and assume a solar abundance for simplicity. We will discuss this treatment in Section~\ref{sec:Discussion}. 

\section{Results} \label{sec:Results}
Equipped by the CSM models described in Section~\ref{Method:CSM}, we next calculate the hydrodynamic evolution of a Type Ib/c SNR up to an age of $10^4$ yr, and the non-thermal emissions resulted from the interaction of the SNR blastwave with the CSM environments.

\subsection{Hydrodynamics}

\begin{figure}[ht!]
\plotone{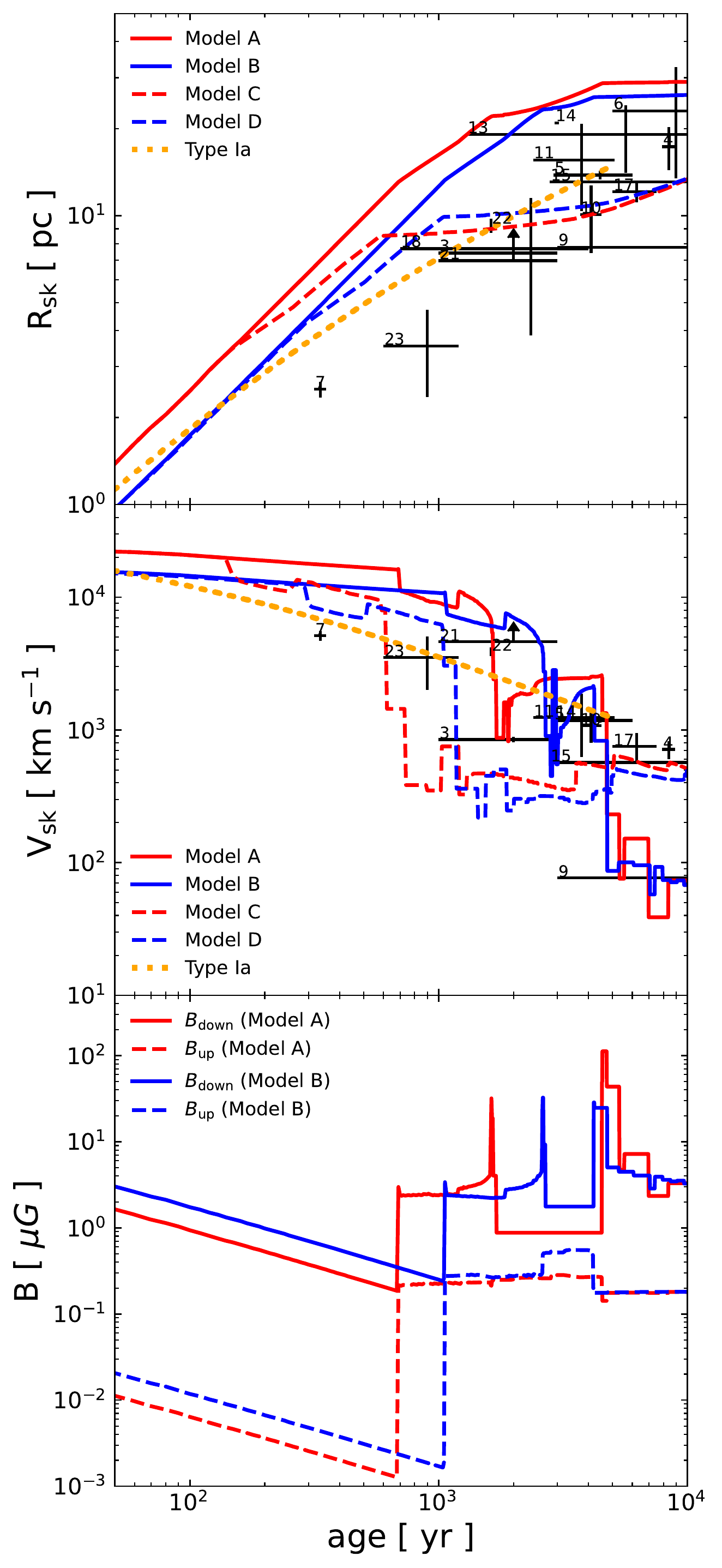}
\caption{The hydrodynamic evolution of a Type Ib/c SNR. The upper panel shows the time evolution of the forward shock radius, the middle panel shows the shock velocity as a function of SNR age, and the lower panel shows the magnetic field at the immediate upstream and downstream of the shock. The line formats in upper and middle panels are the same as in Figure.~\ref{fig:wind}. The orange dotted lines are taken from Model A2 in \citetalias{2019ApJ...876...27Y} for comparison (see text). Actual observational data of selected core-collapse SNRs are overlaid, for which the references can be also found in \citetalias{2019ApJ...876...27Y}. In the lower panel, the red (blue) line shows the magnetic field strengths near the forward shock from Model A (B). The solid (dashed) lines represent values measured at the immediate downstream (upstream) of the forward shock. \label{fig:RV_t}}
\end{figure}

\begin{figure}[ht!]
\gridline{\fig{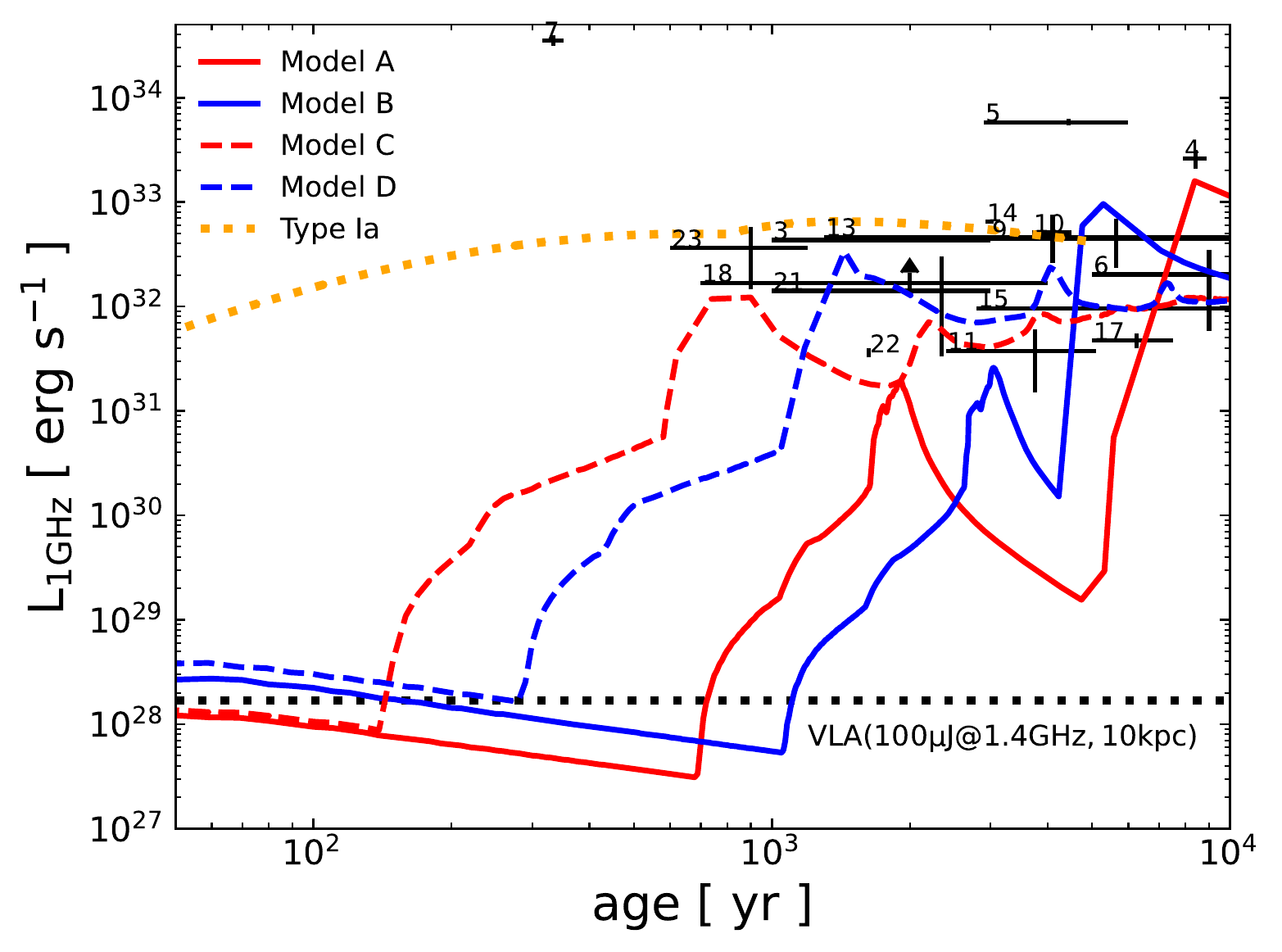}{0.4\textwidth}{(a)}}
\gridline{\fig{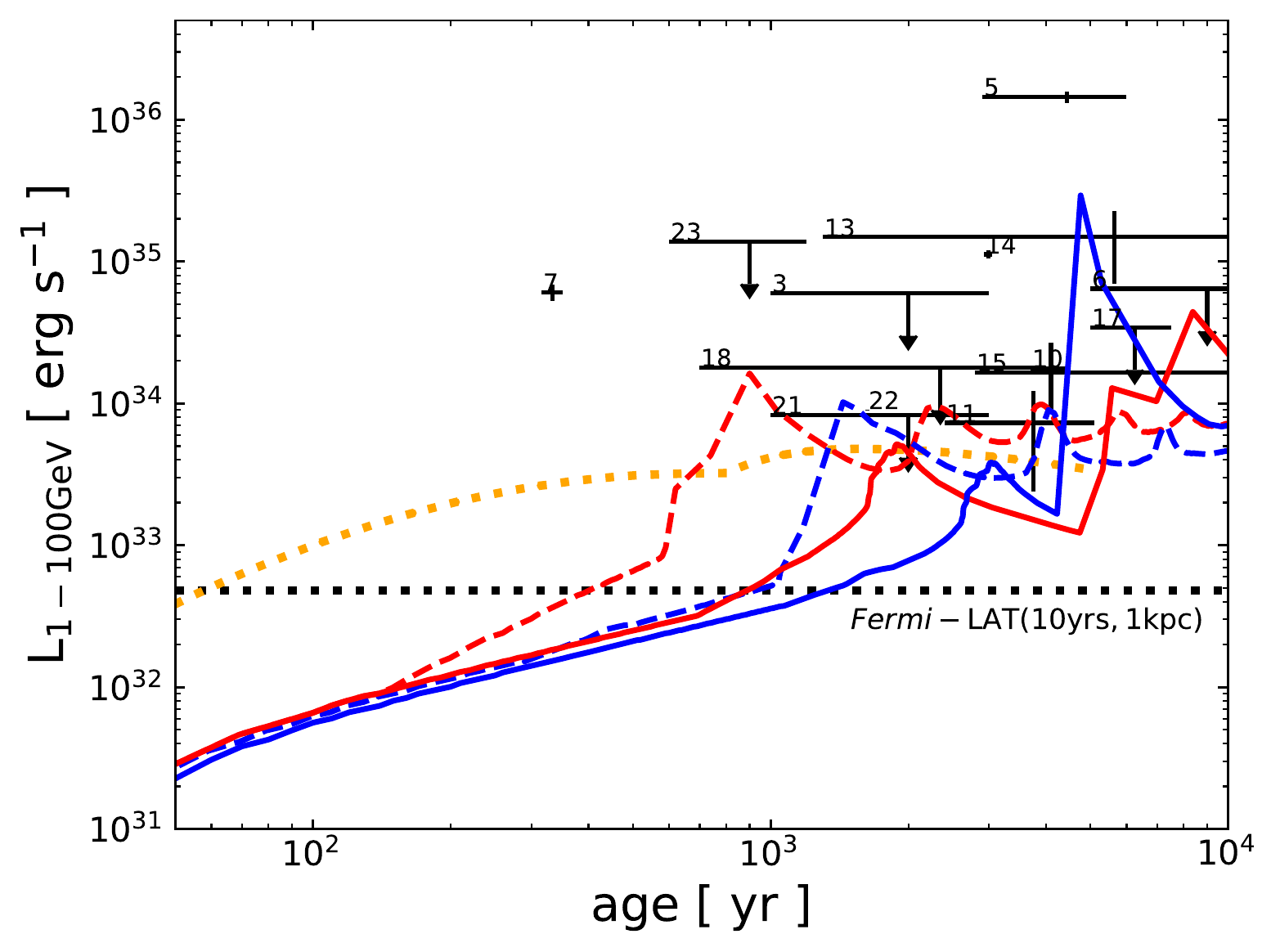}{0.4\textwidth}{(b)}}
\gridline{\fig{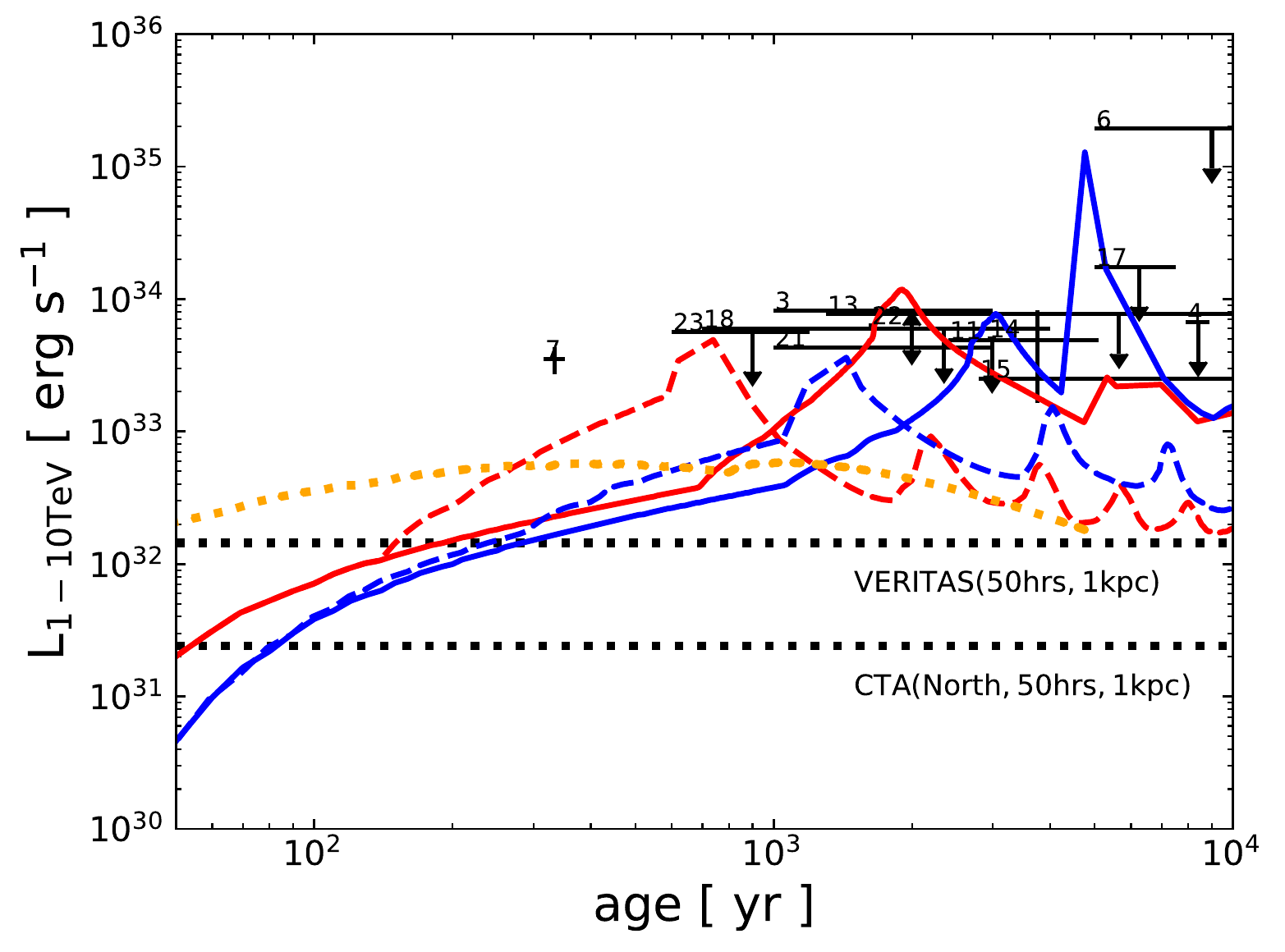}{0.4\textwidth}{(c)}}
\caption{Light curves of the 1~GHz radio continuum (panel (a)), $\gamma$-ray emissions integrated over the 1-100 GeV band (panel (b)) and 1-10 TeV band (panel (c)) are shown. The line formats are the same as in Figure~\ref{fig:RV_t}. The detection limit of VLA in panel (a), {\it Fermi}-LAT in panel (b), and VERITAS and CTA in panel (c) are plotted with black dotted lines, respectively. The results from multi-wavelength observations of selected SNRs as shown in Figure~\ref{fig:RV_t} are also overlaid. \label{fig:lc}}
\end{figure}

\begin{figure}[ht!]
\gridline{\fig{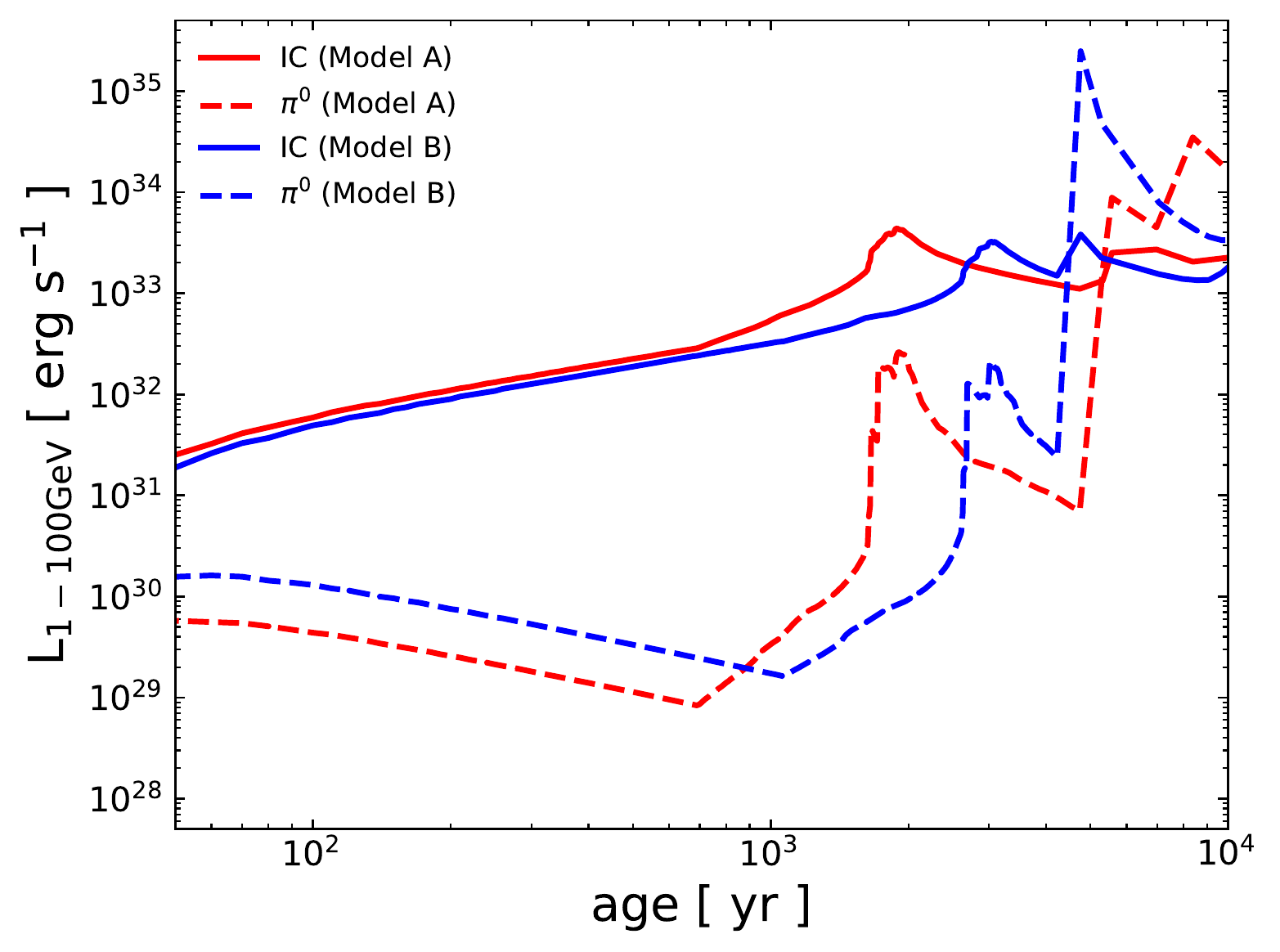}{0.4\textwidth}{(a)}}
\gridline{\fig{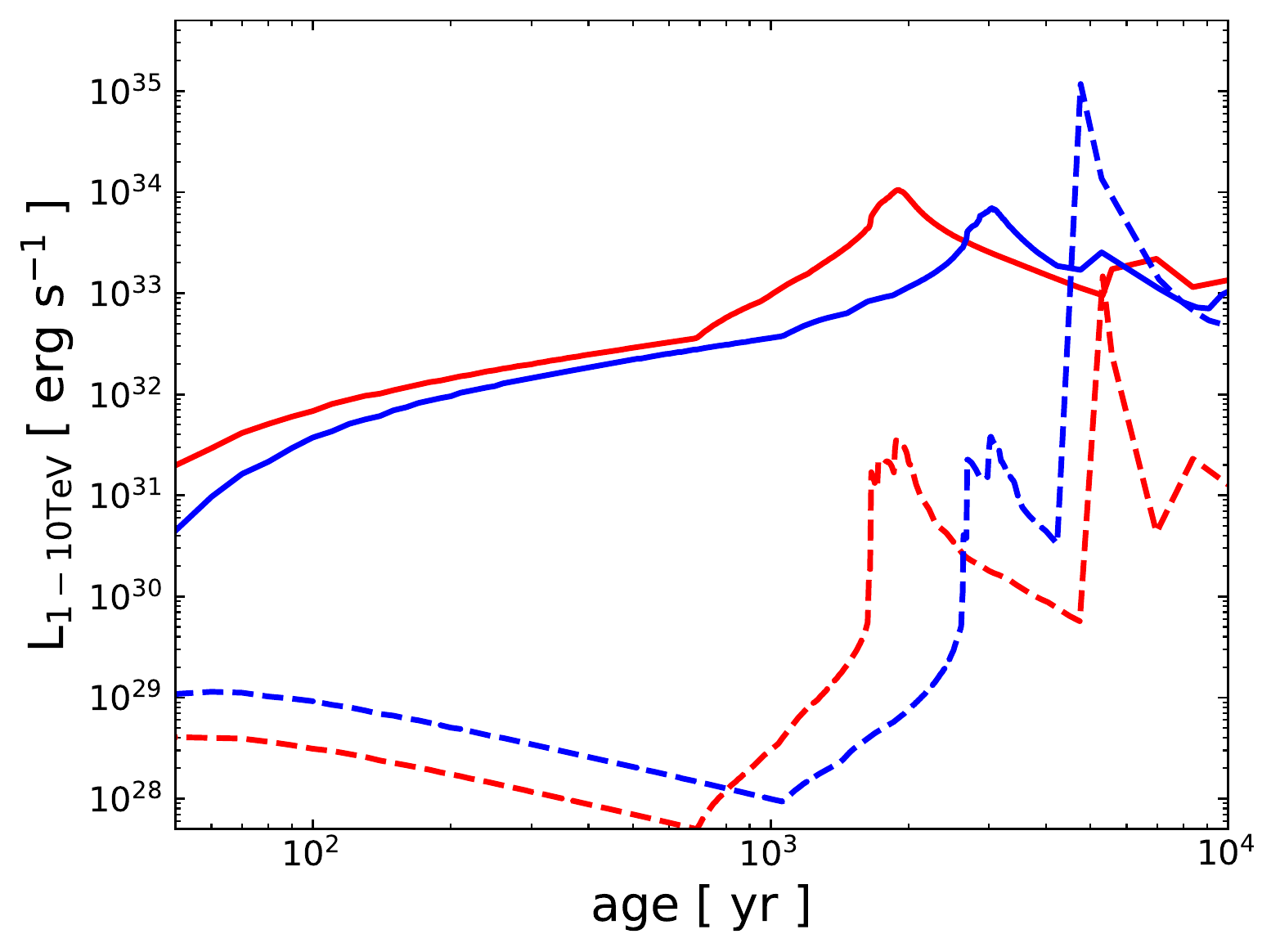}{0.4\textwidth}{(b)}}
\caption{Same as panels (b) and (c) in Figure~\ref{fig:lc}, but the contribution from each emission component is shown separately. The solid (dashed) line shows the IC ($\pi^0$ decay) component, and the red (blue) color represents the results from Model A (B). \label{fig:icpi0}}
\end{figure}

\begin{figure*}[ht!]
\plotone{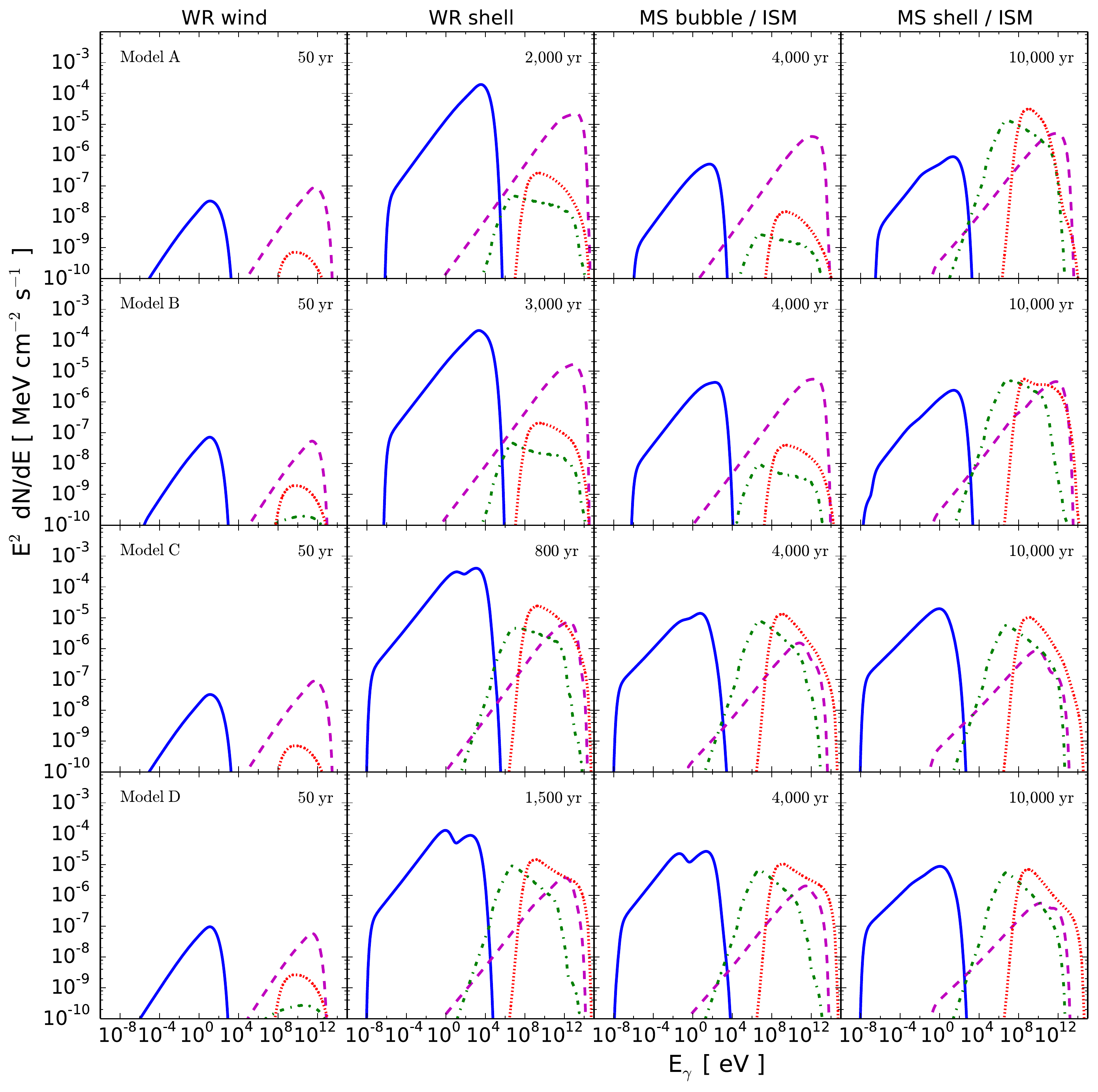}
\caption{Broadband SED from a Type Ib/c SNR with different progenitor masses and CSM models (top to bottom) and at different ages (left to right). The exact age is shown in each panel at the upper right corner. The emission components correspond to synchrotron (blue solid), IC (magenta dashed),  non-thermal bremsstrahlung (green dot-dashed) and $\pi^0$ decay (red dotted). The distance from a source is assumed to be 1~kpc.\label{fig:SED}}
\end{figure*}

Figure~\ref{fig:RV_t} shows the time evolution of the SNR radius $R_\mathrm{sk}$ (upper panel), shock velocity $V_\mathrm{sk}$ (middle panel), and the magnetic field $B(x)$ at the shock position (lower panel) for each model. Similar to \citetalias{2021ApJ...919L..16Y}, we also plot the results from a Type Ia SNR model for comparison (see \citetalias{2019ApJ...876...27Y} and \citetalias{2021ApJ...919L..16Y} for details.). Observational data of selected core-collapse SNRs are also overlaid as black data points. These SNRs are chosen from the $\gamma$-ray source catalog of {\it Fermi} \citep{2016ApJS..224....8A} and H.~E.~S.~S. \citep{2018A&A...612A...3H}, and the exact values and references can be found in \citetalias{2019ApJ...876...27Y}.

We first look at the results of models A (red solid line) and B (blue solid line) in upper and middle panels. In the early phase ($t\le1,000\ \mathrm{yr}$), the SNR forward shock freely expands into the tenuous unshocked WR wind with a velocity $\sim10,000\ \mathrm{km\ s^{-1}}$. Afterwards at an age of ($1,000 \ \mathrm{yr}\le t\le5,000\ \mathrm{yr}$), the SNR blastwave collides with the WR shell, and the shock speed decreases to $\le1,000\ \mathrm{km\ s^{-1}}$. The shock eventually breaks out from the WR shell into the low-density hot MS bubble, and the shock velocity restores to $\sim2,000\ \mathrm{km/s}$. In the late phase ($t\ge5,000\ \mathrm{yr}$), the blastwave  hits a dense wall at the MS shell and decelerates to $\le100\ \mathrm{km\ s^{-1}}$. As the shock sweeps up the large amount of gas contained inside the MS shell, the SNR makes a transition to its radiative phase, and slowly expands into the ISM region after the shock breaks out from the MS shell. The differences between these models are mainly in the timings of transition into each dynamical phase as stated above, which in turn originate from the differences in the mass-loss rates and durations in each pre-SN evolutionary phases, as well as the ejecta mass.

From models C (red dashed line) and D (blue dashed line) in upper and middle panels, we find that the evolution in the early phase is similar to model A and B as the SNR expands into the unshocked WR wind until it hits a the termination shock and starts decelerating at an age $\sim\ 100\ \mathrm{yr}$. Afterwards, the shock collides with the dense WR shell and slows down to a velocity of the order of 100 km s$^{-1}$ at an age of around 1,000 yr. As this happens, the SNR again sweeps up a large amount of gas inside the WR shell and enters the radiative phase. When the radiative shock runs into the ISM region, the expanding hot SN ejecta heated by the reverse shock pushes the cold dense shell formed behind the radiative forward shock outward, causing the forward shock velocity to oscillate \citep[see, e.g.,][]{2015ApJ...806...71L}. The differences between these models are the same as what we have described above. From these results, we can see that the (non-)existence of a MS bubble critically affects the dynamical evolution of a Type Ib/c SNR.

In the lower panel, we can see that the magnetic field strengths at the immediate downstream of the shock (solid lines) are amplified from those in the upstream (dashed lines). Their evolution reflects closely the CSM structure and the shock velocity which are critical parameters for the particle acceleration efficiency and therefore the strength of the CR-driven magnetic turbulence.

\subsection{Non-thermal emissions}
Figure~\ref{fig:lc} shows the light curves of the 1~GHz radio continuum (panel (a)), $\gamma$-ray emissions in the 1-100 GeV band (panel (b)) and the 1-10 TeV band (panel (c))\footnote{We note that X-ray emission is also important for deciphering the properties of SNRs. X-rays from SNRs are produced by not only synchrotron radiation but also thermal components including bremsstrahlung, various continua, and line emission from the hot plasma confined between the forward and reverse shock. By focusing on the non-thermal components in this work, we postpone the presentation of light curves in the X-ray bands to a future work in which a proper implementation of the thermal emission will be included.}. The line formats and colors are the same as in Figure~\ref{fig:RV_t}. The contributions from IC and $\pi^0$ decay are independently plotted in Figure~\ref{fig:icpi0}. The corresponding spectral energy distribution (SED) of each model is plotted in Figure~\ref{fig:SED} at four characteristic ages from left to right, which is explicitly indicated at the upper right corner of each panel. 

From the results of models A and B, we can observe that both the GeV and TeV $\gamma$-ray luminosities gradually increase with time, while the radio counterpart decreases during the first 1,000 yrs. This can be understood as follows. The dominant emission mechanisms for the radio emission is synchrotron radiation, while that for the $\gamma$-rays is IC for both the GeV and TeV bands (see Figure~\ref{fig:icpi0} and the left panels in Figure~\ref{fig:SED}). The synchrotron emissivity is proportional to both the total number of the non-thermal electrons as well as the square of the downstream magnetic field strength. While the wind density drops with radius as $r^{-2}$ so that the injection rate becomes smaller with time, the number of accelerated electrons integrated over the volume of the SNR does increase with time as they are advected and accumulate in the downstream.
The magnetic field strength immediately upstream from the shock is proportional to $r^{-1}$ from the density structure, and the field strength behind the shock further decreases from adiabatic expansion and flux conservation as the shocked gas advects downstream, assuming that the magnetic fields are frozen in the shocked plasma. The overall synchrotron flux hence decreases with the expansion of the SNR. On the other hand, the IC emissivity is proportional to the product of the number of accelerated electrons and the energy density of the target photon field. Because the CMB is assumed as the photon target of IC in this work which is constant in space, IC flux increases with time as the shock keeps accelerating electrons from the inflowing wind material. 

After that, as the shock approaches the WR shell,
the luminosities in both radio and $\gamma$-rays increase with time and reach their first maxima at $\sim$2,000 to 3,000 yrs. Until $\sim$4,000 to 5000 yrs, the SNR expands and breaks out into the tenuous MS bubble, and the non-thermal emission suffers a decay of 1 to 2 orders of magnitude from the rapid adiabatic loss, but this declination does not make the SNR undetectable by current and future instruments because the spatial extend of the MS bubble is compact as we have already explained in Section~\ref{Method:CSM}. Finally, the SNR collides with the dense MS shell 
and becomes bright again in all wavelengths. Once the shock enters the ISM region, it becomes hard for the shock to accelerate particles efficiently anymore due to its low velocity, and the luminosities gradually decrease with time via adiabatic loss again.

As mentioned in Section~\ref{Method:CSM}, the WR shells in model C and D are located at smaller radii than those in model A and B (see Figure~\ref{fig:wind} again), so that the luminosities begin to rise earlier from a few 100 yrs and reach the maximum brightness at around 1,000 yrs. Until 10,000 yrs, the luminosities stay at more-or-less the same level except for slight oscillations originating from 
the velocity fluctuation of the radiative shock as seen in Figure~\ref{fig:RV_t}.

From the SED in Figure~\ref{fig:SED}, we can also see a steepening of the $\pi^0$ decay spectra in all models, which comes from the steepening of underlying proton spectrum. The power-law index of the proton spectrum is roughly obtained as $d\ln{f(p)}/d\ln{p}\sim -3S_\mathrm{tot}/(S_\mathrm{tot}-1)$, where $S_\mathrm{tot}$ is the effective compression ratio \citep[e.g.,][]{CBAV2009} which is determined by the difference between the shock velocity and the velocity of the magnetic scattering centers. In situations where MFA is efficient, the effective compression ratio can become smaller than 4 and the resulting proton spectra hence steepen. As a result, when $\pi^0$ decay is the dominant emission channel in the late evolutionary phase, the total $\gamma$-ray spectrum is characterized by a soft spectrum.

We note that the luminosity of the ``Type Ia" model referenced here is relatively large compared to the Type Ib/c models, especially in the early phase. This is stemming from differences in the assumed DSA injection rate $\chi_\mathrm{inj}$ and the surrounding ISM/CSM environment. The injection rate $\chi_\mathrm{inj}$ determines the amount of particles injected into the acceleration process and the resulting acceleration efficiency \citep[see, e.g.,][]{2004APh....21...45B,2005MNRAS.361..907B}. Here, we have adopted $\chi_\mathrm{inj}=3.6$ for the Type Ia model and $\chi_\mathrm{inj}=3.75$ for the Type Ib/c models based on \citetalias{2019ApJ...876...27Y}. More importantly, according to our CSM models, Type Ib/c SNRs expand into a tenuous wind cavity whereas the Type Ia model adopts a uniform ISM-like environment with $n_\mathrm{ISM}=0.1\ \mathrm{cm}^{-3}$ as in \citetalias{2019ApJ...876...27Y}. While the absolute luminosities do depend on the DSA parameters which should be constrained by observation data of individual SNRs, our results show that the light curve of a Type Ia SNR evolving in a more-or-less uniform ISM is expected to be much flatter and uncharacteristic compared to the remnants of Type Ib/c SNe, for which the latter heavily anchors to the highly inhomogeneous structure of the CSM environment and hence the progenitor mass loss history.

We can now assess the observational detectability of a Type Ib/c SNR based on our models. The sensitivities of various instruments are plotted in all panels in Figure~\ref{fig:lc} with black dotted lines. For the radio band, we compare the detection limit of the Very Large Array (VLA) with our models. The sensitivity for a targeted observation of objects like radio galaxies and active galactic nuclei is $\sim100\ \mu\mathrm{Jy}$ at 1.4 GHz \citep[e.g.][]{2004AJ....128.1974S,2012MNRAS.421.3060S}, and the lower limit from a source at a distance of 10 kpc therefore corresponds to $\sim2\times10^{28}\ \mathrm{erg\ s^{-1}}$. We note that non-targeted sky surveys have much shallower sensitivities, but this does not affect the following discussion and our main conclusion. We use the sensitivity of the {\it Fermi} Large Area Telescope ({\it Fermi}-LAT) for the GeV $\gamma$-rays, and VERITAS (Very Energetic Radiation Imaging Telescope Array System) and the Cherenkov Telescope Array (CTA) for TeV $\gamma$-rays. Based on 10 yrs of survey data\footnote{\url{https://www.slac.stanford.edu/exp/glast/groups/canda/lat_Performance.htm}} \citep[see, for details,][]{2020ApJS..247...33A,2020arXiv200511208B}, the flux sensitivity of {\it Fermi}-LAT in the 1-100 GeV is $\sim 2\times10^{-12}\ \mathrm{erg\ cm^{-2}\ s^{-1}}$, which corresponds to a luminosity $\sim4.8\times10^{32}\ \mathrm{erg\ s^{-1}}$ for a $\gamma$-ray source at 1 kpc. That of VERITAS and the northern telescopes of CTA in the 1-10 TeV band with an observation time of 50 hrs is $\sim 6\times10^{-13}\ \mathrm{erg\ cm^{-2}\ s^{-1}}$ and $\sim 10^{-13}\ \mathrm{erg\ cm^{-2}\ s^{-1}}$, respectively\footnote{VERITAS specifications from \url{https://veritas.sao.arizona.edu/about-veritas/veritas-specifications}, and CTA performance from \url{https://www.cta-observatory.org/science/ctao-performance/}}. For a source at a distance of 1 kpc, the detection limits are $\sim1.4\times10^{32}\ \mathrm{erg\ s^{-1}}$ and $\sim2.4\times10^{31}\ \mathrm{erg\ s^{-1}}$. In this calculation, we do not take other effects like interstellar absorption into account. 

Our results show that Type Ib/c SNRs are most probably too faint to be observed in radio and GeV $\gamma$-rays in the first 1,000 yrs after explosion, although they can potentially be detected as a TeV-bright SNR in the CTA era if they are close by ($\lesssim$ 1 kpc). On the other hand, the SNRs are bright enough to be detectable in all wavelengths from 1,000 to 10,000 yrs after the blastwave has swept through the low-density WR wind and starts to interact with the denser CSM beyond the WR wind. We can conclude that Type Ib/c SNRs are very likely to experience a ``resurrection'' in non-thermal brightness, meaning that they are too dark to be observable in the first 1,000 yrs but re-brighten significantly afterwards until 10,000 yrs. If the MS bubble does not exist, even younger SNRs can become detectable. 
However, we note that while the occurrence of the ``resurrection'' is a robust prediction of our models, its exact timing depends on various additional factors which we have not fully explored in our parameter space, such as the nature of the progenitors including their ZAMS masses and mass-loss rates, and their surrounding ambient environment in which they evolve (e.g., in or near a giant molecular cloud (MC)).

\section{A comparison with Type II SNRs} \label{sec:Discussion}

\subsection{Light curves}

\begin{figure}[ht!]
\gridline{\fig{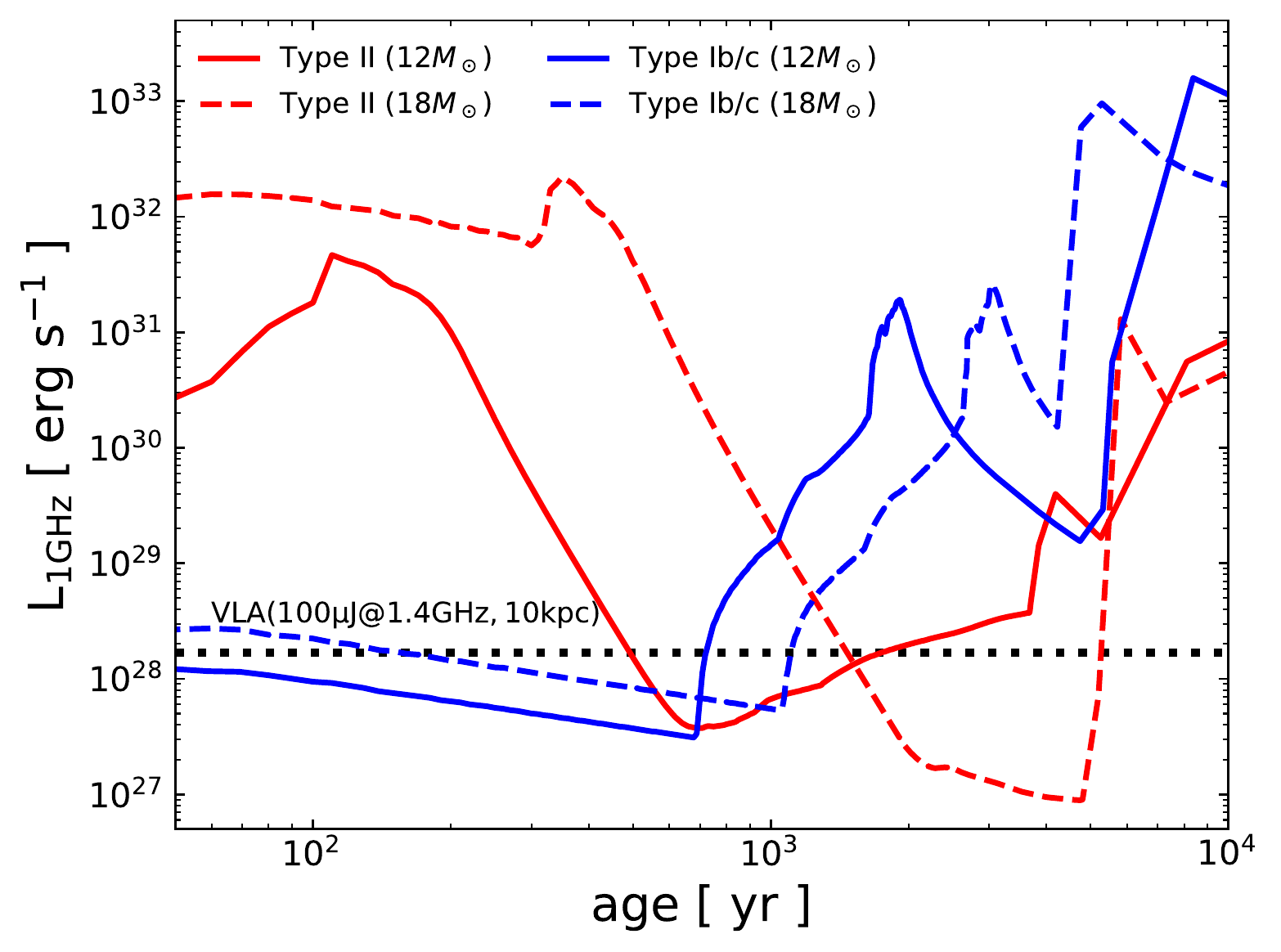}{0.4\textwidth}{(a)}}
\gridline{\fig{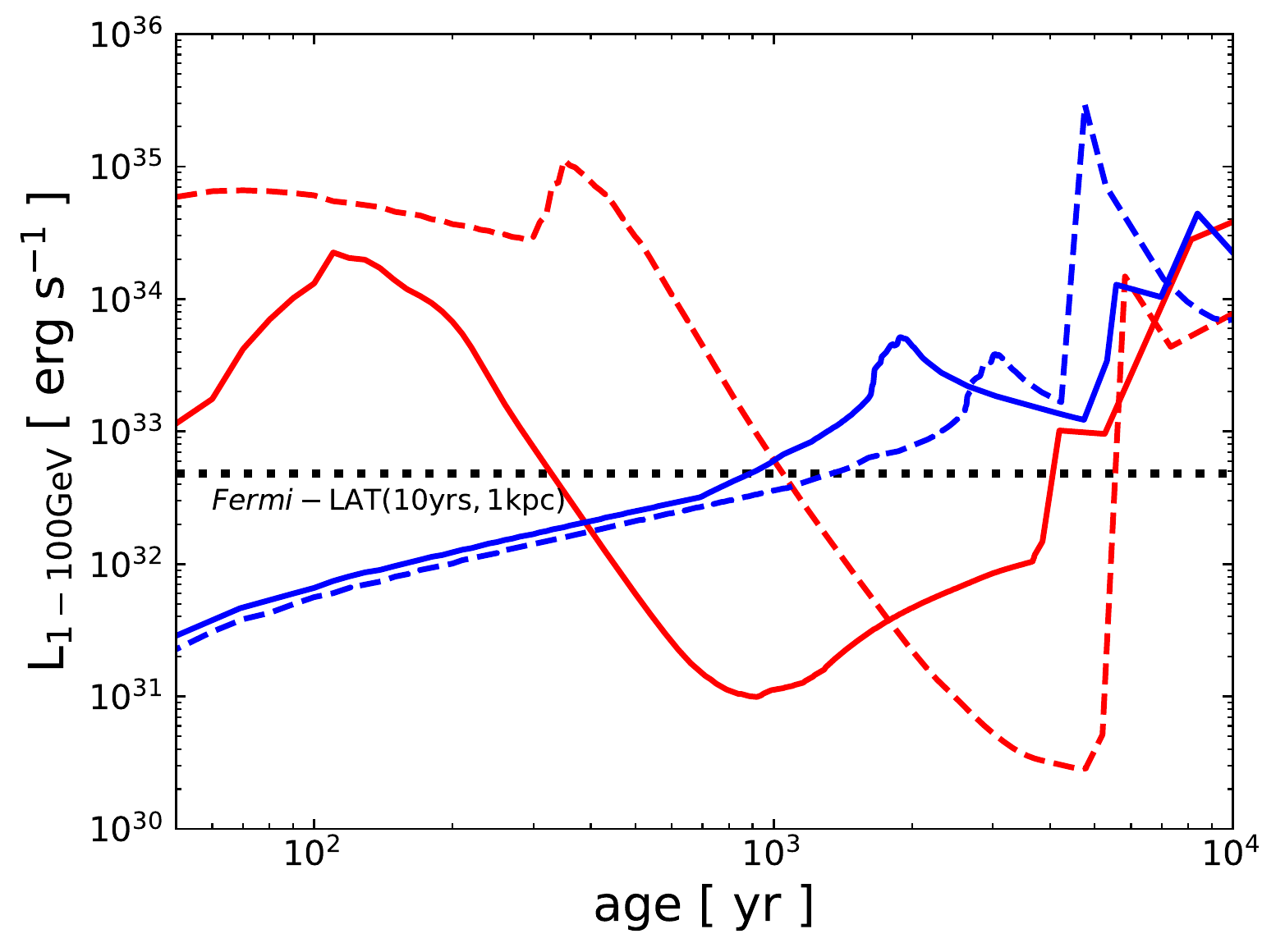}{0.4\textwidth}{(b)}}
\gridline{\fig{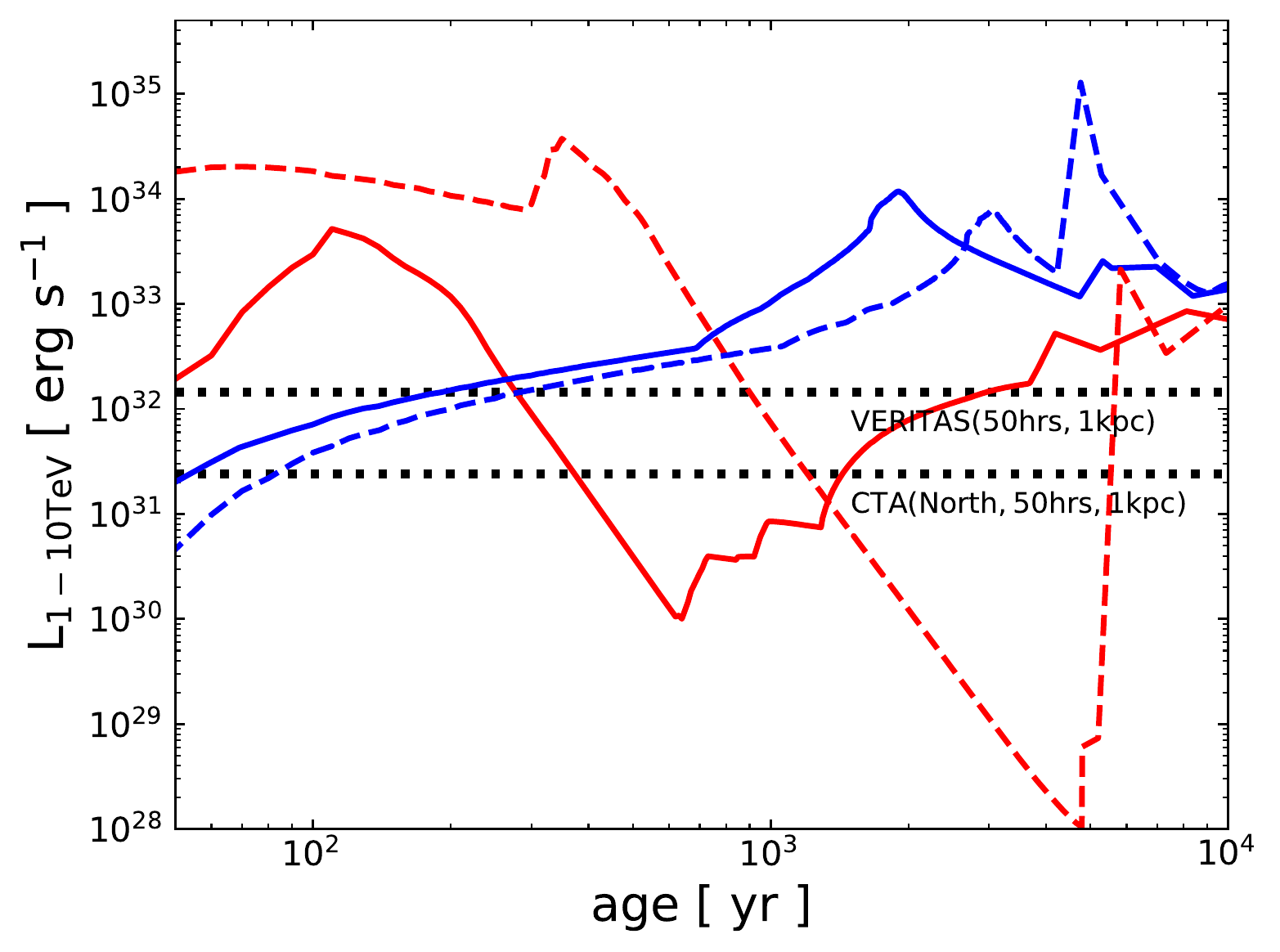}{0.4\textwidth}{(c)}}
\caption{Comparison of Type II SNRs (red lines; \citetalias{2021ApJ...919L..16Y}) and Type Ib/c SNRs (blue lines; this work). Solid lines plot the results of $M_\mathrm{ZAMS}=12\ M_\odot$ model and dashed lines show those of $18\ M_\odot$ model for both colors. The detection limit of various detectors are also plotted with black lines as shown in Figure~\ref{fig:lc}. \label{fig:IIvsIbc}}
\end{figure}

\begin{figure}[ht!]
\gridline{\fig{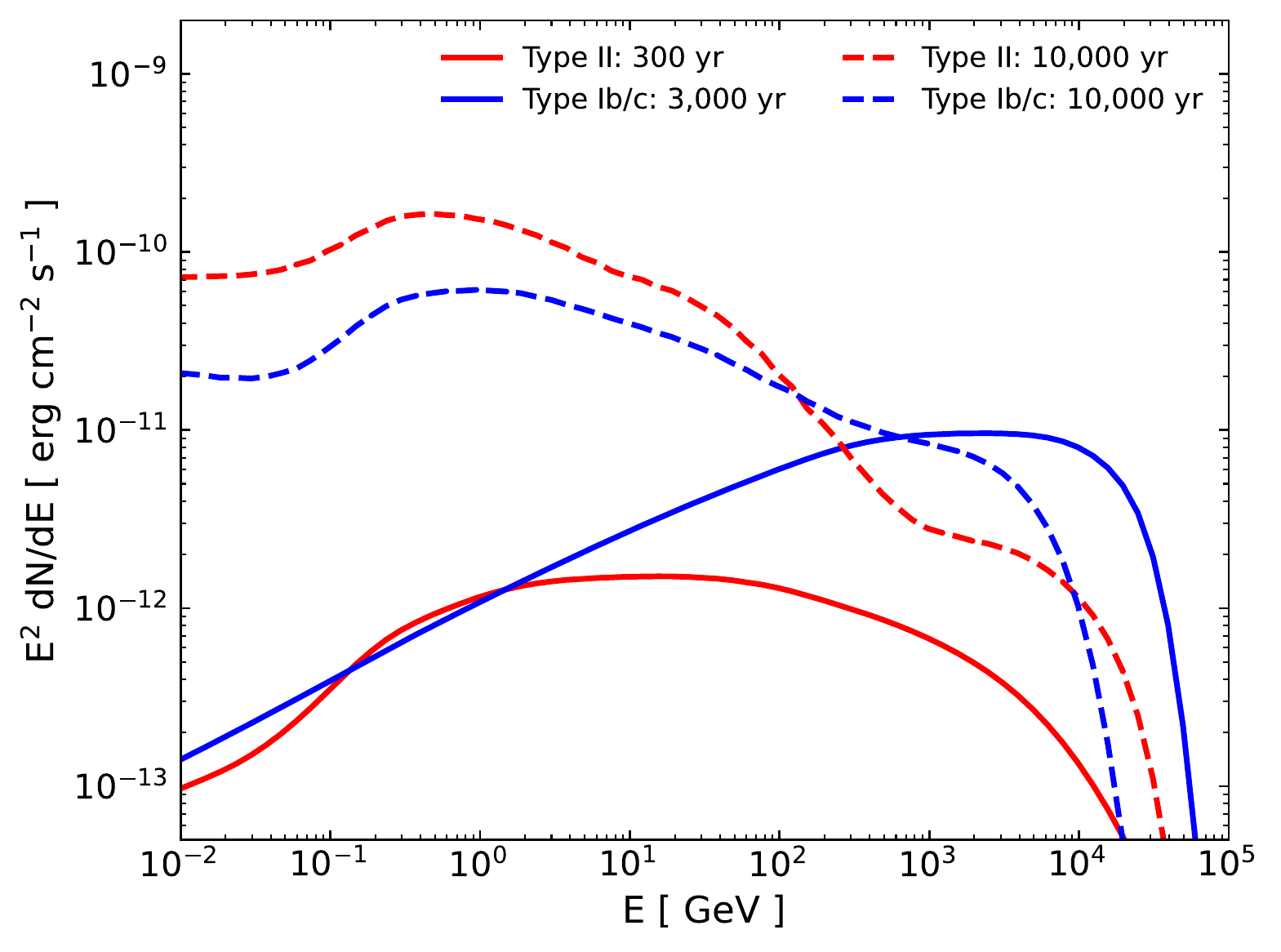}{0.4\textwidth}{(a)}}
\gridline{\fig{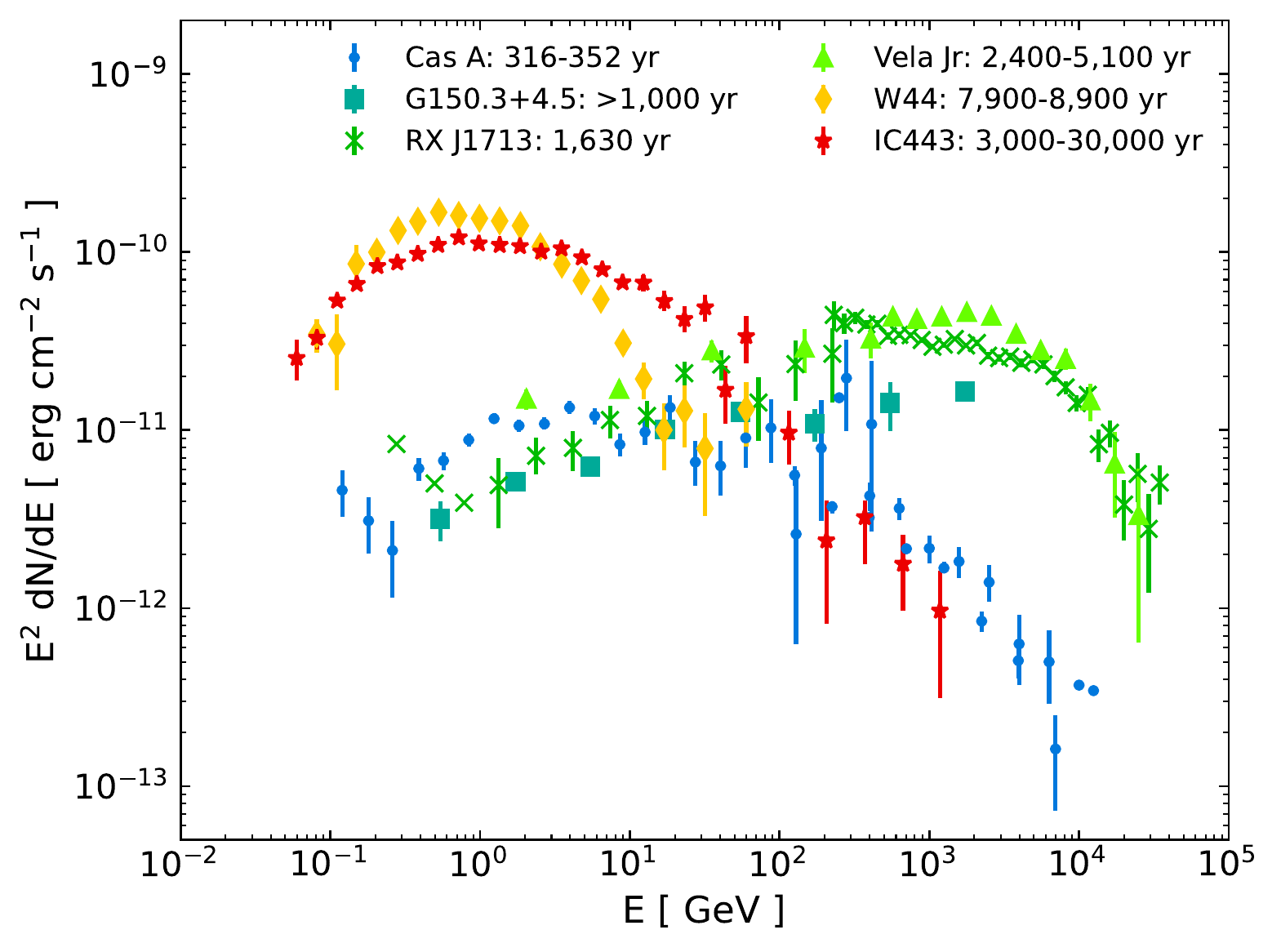}{0.4\textwidth}{(b)}}
\caption{Upper panel (a): simulated $\gamma$-ray SED from this work and \citetalias{2021ApJ...919L..16Y}. Red lines show the results of Model A in \citetalias{2021ApJ...919L..16Y} at 300 yr (solid) and 10,000 yr (dashed), and Blue ones correspond to those of Model A in this work at 3,000 yr (solid) and 10,000 yr (dashed). Lower panel (b): SEDs of several core-collapse SNRs. References of each SNR: Cas A \citep{2017MNRAS.472.2956A,2020ApJ...894...51A}, G150.3+4.5 \citep{2020A&A...643A..28D}, RX J1713 \citep{2018AA...612A...6H}, Vela Jr \citep{2011ApJ...740L..51T,2018AA...612A...7H}, W44 \citep{2013Sci...339..807A}, and IC443 \citep{2007ApJ...664L..87A,2013Sci...339..807A}. The values of age estimations are taken from SNRcat \citep[][and \url{http://www.physics.umanitoba.ca/snr/SNRcat/}]{2012AdSpR..49.1313F}.\label{fig:SNRs}}
\end{figure}

The present results as complemented with the results of \citetalias{2021ApJ...919L..16Y} provides a new picture of SNR evolution highlighted by the difference between Type II and Ib/c SNRs, which is directly linked to their progenitor evolution. Figure~\ref{fig:IIvsIbc} compares the model light curves of Type II SNRs from \citetalias{2021ApJ...919L..16Y} and Type Ib/c SNRs presented in this study. The red lines correspond to Type II remnants (see Models A and B in \citetalias{2021ApJ...919L..16Y}) and the blue lines to Type Ib/c objects (this work). The solid and dashed lines represent models with $M_\mathrm{ZAMS}=12\ M_\odot$ and $18\ M_\odot$, respectively. In \citetalias{2021ApJ...919L..16Y}, the authors proposed that Type II SNRs are very likely to experience a ``dark age'' in which the SNRs become too faint to detect in multi-wavelengths for a prolonged period of time. Meanwhile, this work suggests that Type Ib/c SNRs can experience a ``resurrection'' after a certain age. Combining these results and for a fixed condition of the ambient ISM, it can be found that Type II SNRs tend to be bright when Type Ib/c ones are faint, and vice versa. 
These results suggest a profound implication that there may exist an observational bias in the detected SNR population, in which there is a correlation between the SNR ages and their originating SN types and progenitor natures, thus providing an additional tool for the back-engineering of the observed SNRs and linking them to their progenitor stars and SN explosions.
%
%

However, the determination of SNR age (and distance for that matter) is usually a non-trivial task. Except when a SNR is identified with historical SN like Tycho's and Kepler's SNR, the age is generally obtained by from the sigma-D relation \citep[e.g.,][]{1968AJ.....73...65P,1976MNRAS.174..267C,1998ApJ...504..761C}, which can involve large uncertainties. \cite{2020PASJ...72...72S} shows that the dynamical age from fitting the apparent diameter with the Sedov solution provides good agreement with the plasma age inferred from X-ray observations of the non-equilibrium ionization plasma in several SNRs (see their Figure~9). Further improvements in the accuracy of age determination through multi-wavelength observations are hence critical for linking any observed core-collapse SNR to its progenitor origin and SN type. 

\subsection{Spectral properties}

In addition, we found that our results well reproduce the observations of several $\gamma$-ray bright SNRs as well in terms of their spectral properties. As mentioned above, Type II SNRs are bright in $\gamma$-rays in their early evolutionary phase. As shown by Figure~4 in \citetalias{2021ApJ...919L..16Y}, their dominant $\gamma$-ray emission process is via $\pi^0$ decay. On the other hand, Type Ib/c SNRs are $\gamma$-ray bright in ages when their Type-II counterparts tend to become faint with the primary emission component being IC emission (see upper panels in Figure~\ref{fig:SED}). From these results, one can expect that the $\gamma$-ray spectrum of an SNR is characterized by (i) a flat spectra produced by $\pi^0$ decay at a very young age ($\le$ 1,000 yrs old) as dominated by a Type II origin, (ii) a hard spectrum from IC emission at intermediate ages ($\le$ 5,000 yrs old) with a Type Ib/c origin, and (iii) a soft spectrum at older ages ($\ge$ 10,000 yrs old) independent of SN type (i.e., a mixture of Type II and Ib/c SNRs) as the shock has collided with the dense MS shell and decelerates, with $\pi^0$ decay being the dominant emission component. In panel (a) of Figure~\ref{fig:SNRs}, we plot our simulation results for the SEDs in the sub-GeV to 100 TeV band of Type II SNRs and Type Ib/c ones at the indicated characteristic ages. Meanwhile, panel (b) shows the observed $\gamma$-ray SEDs of a few core-collapse SNRs. Indeed, we can see that very young objects like Cas A do exhibit a relatively flat spectrum \citep{2010ApJ...714..163A,2013ApJ...779..117Y,2017MNRAS.472.2956A,2020ApJ...894...51A}, while SNRs of a few 1,000 yrs old like RX J1713.7-3946 \citep{2011ApJ...734...28A,2018AA...612A...6H}, Vela Jr \citep{2011ApJ...740L..51T,2018AA...612A...7H}, and G150.3+4.5 \citep{2020A&A...643A..28D} show harder spectra, and the more evolved middle-aged SNRs like IC 443 and W44 \citep{2013Sci...339..807A} typically show very soft spectra. While this general agreement does not necessarily imply that the picture above is applicable for every single individual object, our results imply in general a strong correlation of the $\gamma$-ray spectral properties of a SNR with its progenitor nature and hence the mass-loss history and CSM structure.   

We emphasize that this transition of the SED properties is expected only when we consider the mass-loss histories of the progenitors. For example, \citetalias{2019ApJ...876...27Y} also attempted to calculate the time evolution of core-collapse SNRs with a method similar to this work, but they assumed that the SNRs are embedded within a simple power-law CSM ($\rho \propto r^{-2}$) without considering the pre-SN mass loss history. 
Their more simplistic models did not predict such a SED transition described above regardless of  the choice of parameters such as the mass-loss rate, whereas the transition emerges naturally in this work and \citetalias{2021ApJ...919L..16Y} by using more self-consistent CSM models linked to the evolution of the progenitor stars. Moreover, a mixture of SN types is found to play an important role as well.

\subsection{Additional remarks}

One caveat is that our simulation is one-dimensional and we assume that the SN progenitors evolve into an ISM with a fixed density of $1\ \mathrm{cm}^{-3}$. 
However, many core-collapse SNRs are known to have asymmetrical morphologies, and some of them are known to be interacting with high-density materials like MCs as mentioned above. Hence, we do not expect that our results can be applied to explain detailed properties of every individual SNR. The investigation of multi-dimensional effects and the diversity of the surrounding ISM is postponed to a future work. Nonetheless, we believe that our one-dimensional but sophisticated evolution models succeed to capture a big picture of how the pre-SN evolution of the progenitors which spans millions of years can be linked to the observational properties (non-thermal emission in particular for this work) of their SNRs thousands of years after the explosion.

Another additional factor that we have not explored yet is the effect of a non-solar metallicity. We assume a solar abundance throughout our simulation box, while we expect that the WR wind should possess a metal-rich composition such as Helium and/or Carbon-Oxygen. However, we note that this does not affect our main results because the dominant non-thermal emission mechanisms while the blastwave is inside the WR wind are from synchrotron in radio and IC in GeV-TeV band respectively, for which the effect from an altered metallicity is mainly on the free electron number density. We can estimate that the change of the density in a helium-rich environment is only about a factor different from that in an environment with a solar-like abundance. This error is much smaller than the uncertainties from poorly constrained parameters such as the mass-loss properties, distance and so on. We hence ignore the metallicity for our calculations of non-thermal emissions here as a secondary effect. A future follow-up study will include other processes such as heavy ion acceleration and escape in the stellar wind as well \citep{2010ApJ...725..184B,2011ApJ...729L..13O,2015PhRvL.114q1103A,2015PhRvL.115u1101A,2017PhRvL.119y1101A} and investigate their effects on the resultant non-thermal emission properties.


Finally, it is illustrative to discuss other possible sub-types of SNRs beyond what we have modeled so far. For example, almost 10$\%$ of all core-collapse SNe are classified as Type IIb \citep{2011MNRAS.412.1522S}, of which the representative remnant objects include the Galactic SNR Cassiopeia A \citep{1996ApJ...466..866B,2008Sci...320.1195K}
Their progenitors are believed to be helium stars embraced by a thin hydrogen envelope that is not completely stripped off by the binary interaction and stellar winds. 
They show a diversity in the CSM density, but it is generally larger than the WR wind case and close to the RSG wind. As such, we expect that Type IIb SNRs will evolve in a similar manner with Type II SNRs. In addition, the diversity in the CSM density and thus in the final mass-loss rate is suggested to be linked to the timing of the binary interaction \citep{2015ApJ...807...35M}, which may reflect expected diversity in the initial binary configuration leading to SNe IIb \citep{2017ApJ...840...90O}. Therefore, theoretical investigation adopting realistic mass-loss history for SNe IIb will provide an interesting possibility to further constrain the details of the stellar evolution scenarios and roles of the binary interaction toward SNe. An expansion of our work to model other possible types of SNRs will be found in a follow-up paper.

\section{Conclusion} \label{sec:Conclusion}
Non-thermal emission from various types of SNRs is an effective probe of their surrounding environment and hence the nature and evolution of their progenitor stars. Following the method of \citetalias{2021ApJ...919L..16Y} who focused on Type II SNRs, we have conducted simulations of the long-term evolution of Type Ib/c SNRs interacting with their CSM in this work, taking into account the mass-loss history of their progenitors. The non-thermal emissions produced by the interactions between the accelerated CRs and the surrounding environment are presented.

We show that the non-thermal emissions from Type Ib/c SNRs are faint and below the sensitivities of current and near-future detectors in the early phase when the SNR blastwave is inside the unshoced WR wind region ($t\le1,000\ \mathrm{yr}$), except if the source is extremely close ($d\le1\ \mathrm{kpc}$) and the TeV emission can be potentially picked up by future observatories such as CTA. These objects are also predicted to be non-thermally bright after the SNR shock has begun to penetrate through the WR wind shell ($t\ge1,000\ \mathrm{yr}$). As the SNR shock passes through the dense shell at around 2,000-3,000 yrs, the brightness of SNRs decreases gradually due to the weakening of the shock and fast adiabatic cooling in the hot compact MS bubble until 5,000 yr. Finally, they collide with the dense MS shell and re-brighten again, but gradually lose their punches once more because of the rapid deceleration of the shock into the radiative phase. We conclude that the non-thermal emission from most Type Ib/c SNRs should experience a ``resurrection'' at some point ($\sim1,000\ \mathrm{yr}$ for a typical ambient ISM density of $n = 1$~cm$^{-3}$) for progenitors with ZAMS mass $M_\mathrm{ZAMS}\le18\ M_\odot$. 
While the exact values of the timescales mentioned above are dependent on the (non-)existence of the MS bubble, the ejecta mass, the wind and ISM properties and so on, our conclusion on the predicted general evolution of a Type Ib/c SNR stays robust because it is independent on any fine-tuning of parameters. 

We have also compared the results in this work to a previous study on Type II SNRs as reported in \citetalias{2021ApJ...919L..16Y}. We show that while Type II SNRs are expected to be bright in the first 1,000 yrs or so but faint afterwards for a few 1,000 yrs in both radio and $\gamma$-rays, Type Ib/c SNRs are showing an opposite evolution characteristics, i.e., they are predicted to be dark in the early phase but re-brighten after an age of  about 1,000 yr, assuming an $n = 1\ \mathrm{cm}^{-3}$ ISM density. This contrasting behavior leads to an evolutionary picture for the SNR population which is found to be compatible with the $\gamma$-ray observation of core-collapse SNRs, in particular the observed broadband spectral properties against the SNR ages, which cannot be reproduced by simplistic models without considering the mass loss histories of the SN progenitors. Another profound implication from our results is that there is a possible observational bias in the current and future SNR observations, i.e., the SN type and progenitor origin of an observed SNR are correlated to its age or evolutionary phase. By the inclusion of other SN sub-types (probably including the different kinds of Ia's as well), and as further observational constraints becoming available in the future, we plan to expand our work to provide a more complete description of the SNR population as a whole.   

\begin{acknowledgments}
H.Y. acknowledges support by JSPS Fellows grant No. JP20J10300. S.H.L. acknowledges support by JSPS grant No. JP19K03913 and the World Premier International Research Center Initiative (WPI), MEXT, Japan. K.M. acknowledges support from JSPS KAKENHI grants JP18H05223, JP20H04737, and JP20H00174.
\end{acknowledgments}

\appendix
\restartappendixnumbering
\section{Additional models \label{Appendix}}

\begin{figure}[ht!]
\plotone{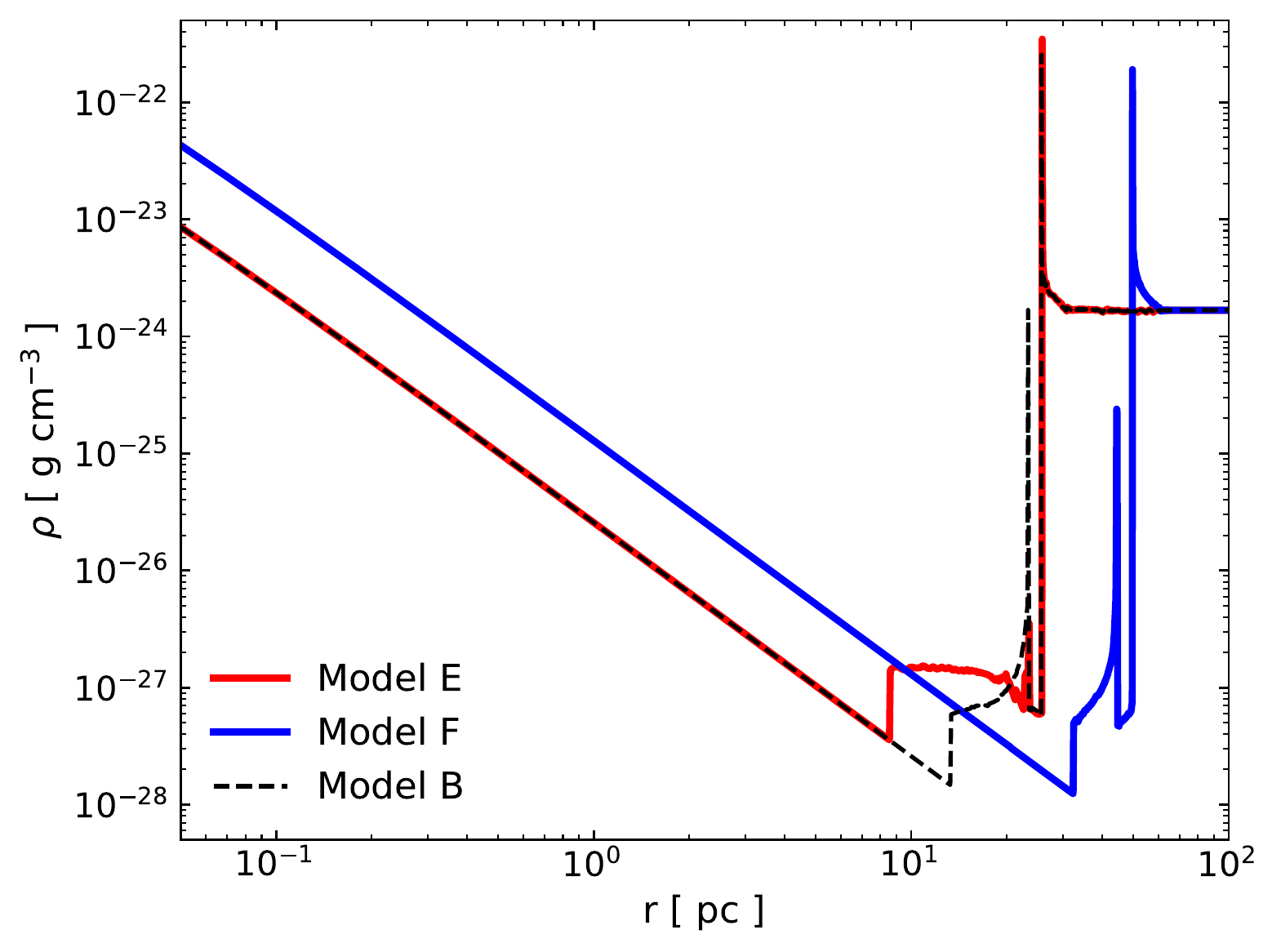}
\caption{Radial density profiles of the CSM for the two additional models. The red solid line corresponds to model E ($\beta_\mathrm{acc}=1$), and the blue solid line to model F ($M_\mathrm{ZAMS}=30\ M_\odot$). The black dashed line shows the result of model B for comparison. \label{App_fig:wind}}
\end{figure}

\begin{figure}[ht!]
\gridline{\fig{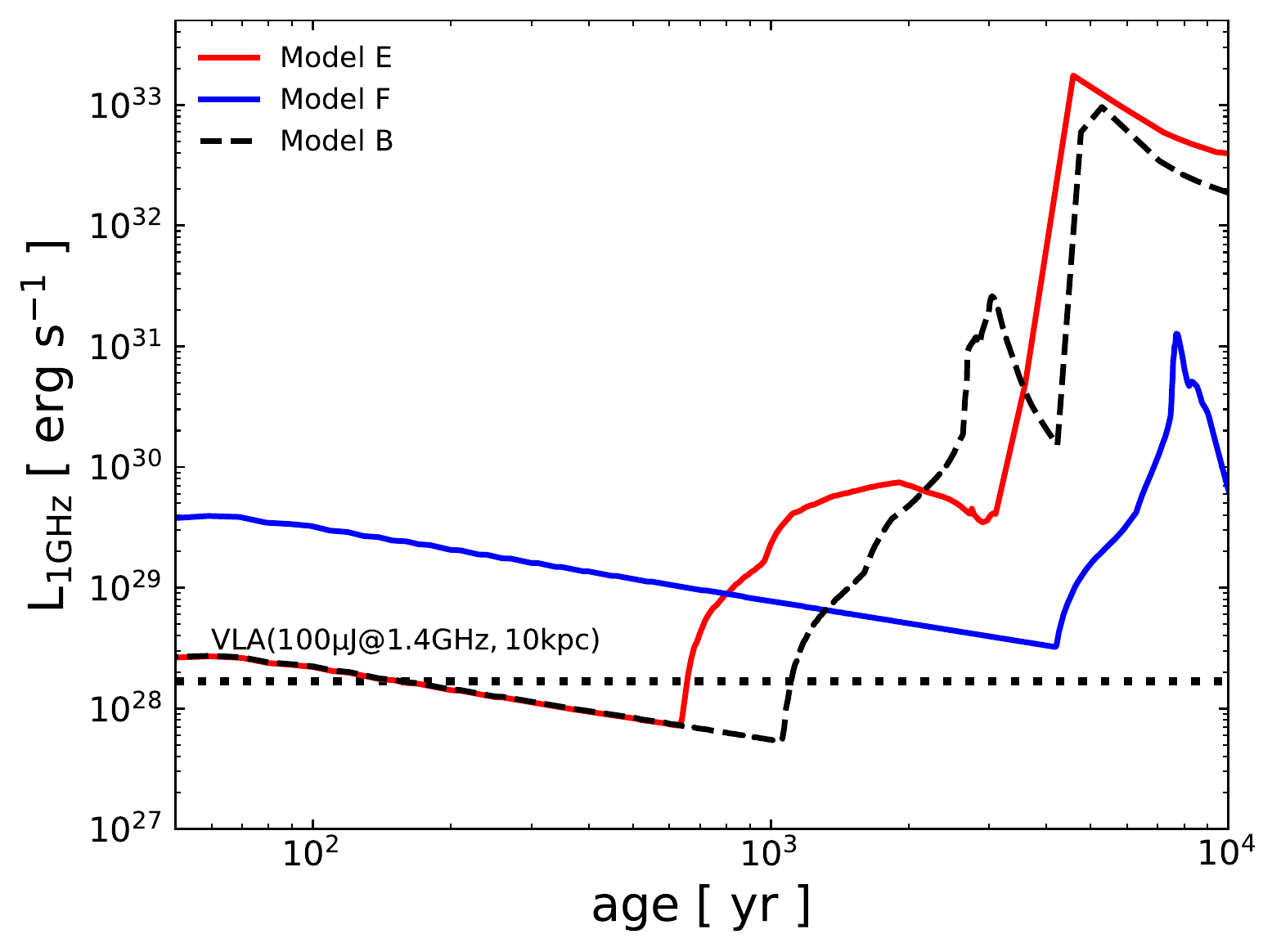}{0.4\textwidth}{(a)}}
\gridline{\fig{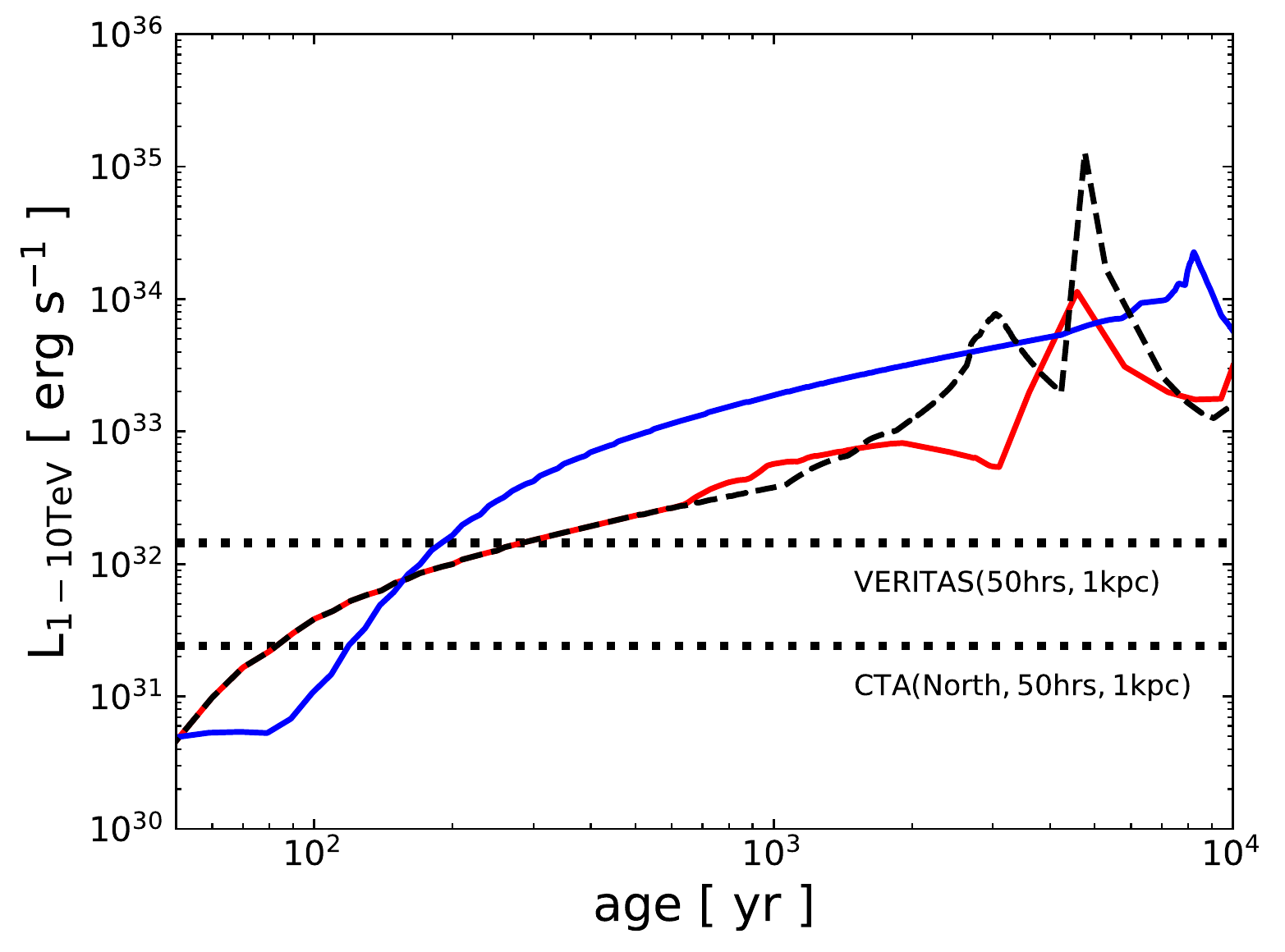}{0.4\textwidth}{(b)}}
\caption{Light curves in 1~GHz radio continuum (upper panel (a)) and 1-10 TeV band (lower panel (b)) for the additional models. The line formats are the same as in Figure~\ref{App_fig:wind}, and the detection limits as shown in Figure~\ref{fig:lc} are also shown. \label{App_fig:lc}}
\end{figure}

As mentioned in Section~\ref{Method:CSM}, we have included two additional models here for reference: (i) a model in which all the matters stripped off by RLOF is accreted onto the secondary star (i.e., $\beta_\mathrm{acc}=1$; hereafter model E), and (ii) a model in which a massive star with $M_\mathrm{ZAMS}=30\ M_\odot$ evolves as a single star without binary interactions (hereafter model F). Figure~\ref{App_fig:wind} shows the CSM density profiles for the two models, and the model parameters are summarized in Table~\ref{App_table:wind}. Figure~\ref{App_fig:lc} plots the time evolution of their non-thermal radio and $\gamma$-ray luminosities. The red and blue solid lines correspond to models E and F respectively, and we also over-plot the results of the fiducial model B as a comparison.

\begin{deluxetable*}{ccccccccc}
\tablecaption{Parameters of two additional models\label{App_table:wind}}
\tablewidth{0pt}
\tablehead{
\colhead{Model} & \colhead{$M_\mathrm{ZAMS}$} & \colhead{Wind Phases} & \colhead{$\dot{M}$} & \colhead{$V_\mathrm{w}$} & \colhead{$M_\mathrm{w}$} & \colhead{$\tau_\mathrm{phase}$} & \colhead{$M_\mathrm{ej}$}\\
\colhead{} & \colhead{($M_\odot$)} & \colhead{} & \colhead{($M_\odot\ \mathrm{yr^{-1}}$)} & \colhead{($\mathrm{km\ s^{-1}}$)} & \colhead{($M_\odot$)} & \colhead{($\mathrm{yr}$)} & \colhead{($M_\odot$)}}
\startdata
E & 18 & MS    & $6.0\times10^{-7}$ & 2000   & 0.3 & $5.0\times10^6$ &     \\
  &    & RLOF  & $0.0$ & 10    & 0.0 & $1.0\times10^4$ &     \\
  &    & WR    & $1.0\times10^{-5}$ & 2000   & 1.0 & $1.0\times10^5$ & 2.5 \\
\hline
F & 30 & MS    & $5.0\times10^{-7}$ & 2000   & 2.0 & $4.0\times10^6$ &     \\
  &    & RSG   & $5.0\times10^{-5}$ & 10     & 18.0 & $3.6\times10^5$ &     \\
  &    & WR    & $5.0\times10^{-5}$ & 2000   & 5.0 & $1.0\times10^5$ & 3.5 \\
\enddata
\tablecomments{The wind parameters and ejecta properties of two additional models. Other details are as written in the footnote of Table~\ref{table:wind}.}
\end{deluxetable*}

\subsection{Model E}
From Figure~\ref{App_fig:wind}, we can see that there are two major differences between models E and B. One is that the density in the WR shell is smaller by more than two orders of magnitude in model E; the other is that the termination shock is sitting at a more inner region in the WR wind. These are caused by their differences in the CSM formation history. As the matters stripped off from the progenitor by RLOF in model E are all accreting onto the secondary star without contributing to the CSM gas distribution, the mass swept by the subsequent WR wind is much smaller than in model B, and the WR shell contains a much smaller mass. This also leads to a faster expansion of the WR wind towards the outlying ISM, resulting into the termination shock propagating further inward against the outgoing unshocked WR wind in the last $10^5$~yrs. 

These differences in the CSM structure are directly reflected in the light curves shown in Figure~\ref{App_fig:lc}. In model E, the SNR shock collides with the termination shock at an earlier time, and the luminosities in both energy bands start to rise from $\sim600\ \mathrm{yr}$. The luminosities reach their maximum values at $\sim2,000\ \mathrm{yr}$, but are smaller than in model B because the mass inside the WR shell is much smaller. This implies that if $\beta_\mathrm{acc}\simeq 1$, it becomes more difficult to detect Type Ib/c SNRs, especially in $\gamma$-rays.  


\subsection{Model F}
One of the distinctive features of model F is a higher mass-loss rate in each wind phase as shown in Table~\ref{App_table:wind}, which is reflected by the higher density in the wind in Figure~\ref{App_fig:wind}. Another difference is that the main mass-stripping mechanism is not via binary interaction but the RSG wind. Nevertheless, the CSM structure of models B and F are found to be qualitatively similar to each other except that it is more spread out in radius for model F. As explained in Section~\ref{Method:CSM}, the high-velocity wind from the WR star sweeps the RSG wind up quickly, forming a similar CSM structure as that in model B. The WR wind in model F has a high ram pressure from the higher mass-loss rate, therefore the wind shell is formed at a more outer region. 

From Figure~\ref{App_fig:lc}, we find that this type of SNRs expand into the high density wind region for about 4,000 yrs, and they are brighter in both radio and $\gamma$-rays than SNRs with a lower mass progenitor in binaries  for the first 1,000 yrs or so. They however becomes relatively faint in radio afterwards during 1,000-4,000 yr due to the fact that the SNR shock is still interacting with the unshocked power-law WR wind, while in model B it has already collided with the dense WR shell. From this point on, the light curves exhibit a similar behavior as model B. This result suggests that a detection of bright non-thermal emission from a very young Type Ib/c SNR (e.g., a couple $100$ yrs old) may imply a high-mass single star for the progenitor, but most probably it will be difficult to distinguish between a single star and binary origin solely from the observed non-thermal emission properties because the details in the pre-SN mass loss history are almost washed away in the emergent CSM structure by the fast WR wind prior to explosion. 


\bibliography{reference}{}
\bibliographystyle{aasjournal}



\end{document}